\documentclass[preprint,12pt]{elsarticle}

\usepackage{amssymb}
\usepackage{amsmath}
\usepackage{bm}

\usepackage{float}
\floatstyle{plaintop}
\restylefloat{table}
\usepackage{caption}
\usepackage{tablefootnote}
\usepackage{threeparttable}
\usepackage{adjustbox}
\usepackage{booktabs}
\usepackage{multirow}
\usepackage{cleveref}
\usepackage{nomencl}

\usepackage{lineno}
\makenomenclature
\makenomenclature

\journal{Advances in Engineering Software}

\begin{document}

\makeatletter
\def\ps@pprintTitle{%
	\let\@oddhead\@empty
	\let\@evenhead\@empty
	\def\@oddfoot{\reset@font\hfil\thepage\hfil}
	\let\@evenfoot\@oddfoot
}
\makeatother

\begin{frontmatter}



\title{A novel, finite-element-based framework for sparse data solution reconstruction and multiple choices}


\author[label1]{Wiera Bielajewa}
\ead{wiera.bielajewa@swansea.ac.uk}

\author[label2]{Michelle Baxter (née Tindall)}
\ead{michelle.baxter@ukaea.uk}

\author[label1]{Perumal Nithiarasu\corref{cor1}}
\ead{p.nithiarasu@swansea.ac.uk}

\affiliation[label1]{organization={Zienkiewicz Institute for Modelling, Data and AI},%Department and Organization
            addressline={Swansea University, Bay Campus}, 
            city={Swansea},
            postcode={SA1 8EN},
            country={United Kingdom (UK)}}

\affiliation[label2]{organization={Culham Science Centre, United Kingdom Atomic Energy Authority (UKAEA)},%Department and Organization 
            city={Abingdon},
            postcode={OX14 3DB},
            country={United Kingdom (UK)}}

\cortext[cor1]{Corresponding author}

\begin{abstract}

Digital twinning is gaining widespread popularity across various areas of engineering, and indeed it offers a capability of effective real-time monitoring and control, which are vital for cost-intensive experimental facilities, particularly the ones where extreme conditions exist. Sparse experimental measurements collected by various diagnostic sensors are usually the only source of information available during the course of a physical experiment. Consequently, in order to enable monitoring and control of the experiment (digital twinning), the ability to perform inverse analysis, facilitating the full field solution reconstruction from the sparse experimental data in real time, is crucial. Such solution reconstruction might be necessary to control a system, if a parameter to be controlled cannot be directly derived from the sparse measurements alone, as oftentimes is the case, for instance maximum temperature within a test piece.

This paper shows for the first time that it is possible to directly solve inverse problems, such as solution reconstruction, where some or all boundary conditions (BCs) are unknown, by purely using a finite-element (FE) approach, without needing to employ any traditional inverse analysis techniques or any machine learning models, as is normally done in the field. This novel and efficient FE-based inverse analysis framework employs a conventional FE discretisation, splits the loading vector into two parts corresponding to the known and unknown BCs, and then defines a loss function based on that split. In spite of the loading vector split, the loss function preserves the element connectivity. This function is minimised using a gradient-based optimisation; and the near real-time operation for heat conduction in a stainless steel plate is achieved.

Furthermore, this paper presents a novel modification of the aforementioned approach, which allows it to generate a range of different solutions satisfying given requirements in a controlled manner. Controlled multiple solution generation in the context of inverse problems and their intrinsic ill-posedness is a novel notion, which has not been explored before. This is done in order to potentially introduce the capability of semi-autonomous system control with intermittent human intervention to the workflow. Having access to a variety of feasible alternatives during the experiment can augment the human decision-making process and assist the operator in evaluating and selecting the most suitable course of action.

\end{abstract}

\begin{keyword}

Solution reconstruction \sep inverse analysis \sep multiple choices \sep sparse data \sep digital twinning \sep finite element method



\end{keyword}

\end{frontmatter}


% \linenumbers
\renewcommand\linenumberfont{\normalfont\small\sffamily}



%%       \citet{<label>} ==> Jones et al. [21]
%%       \citep{<label>} ==> [21]
%%

%\printnomenclature

\section{Introduction}\label{intro}

\printnomenclature
Digital twins and digital twinning are becoming increasingly popular in various engineering fields, which is a trend driven by the desire to have a dynamic virtual representation of a physical system in order to extract as much useful real-time information as possible from the physical asset. A digital twin is a dynamic virtual representation of a physical system, which is continuously updated from real-time data in order to fully and accurately simulate the current behaviour of its physical counterpart \cite{DT_review}. Whereas, digital twinning refers to the process of creating and maintaining digital twins, which becomes important when the dynamic control of the physical system is introduced. When applied to the experimental facilities, digital twinning broadly aims to enhance the sparse experimental measurements collected from the various diagnostic sensors, i.e. provide system monitoring. It also aims to ensure the ability to reach the required experimental conditions within the optimal time frame, i.e. maintain system control. Therefore, the first step towards having a fully-fledged digital twinning process capable of fulfilling these two objectives involves developing an efficient inverse analysis framework, which would be able to directly integrate the experimental data into the simulation in real time in order to obtain a full solution. Full solution might be necessary to achieve desired system control, in cases when a parameter to be controlled cannot be directly derived from the sparse measurements alone, for example maximum temperature within a test piece. This paper presents an inverse analysis framework based on the Finite Element Method (FEM) as a compelling alternative to existing approaches.

In general, a forward problem can be defined as using the applicable model of a system to ascertain the effects of the given causes. Specifically, for the transient problems, the boundary conditions (BCs), the initial conditions (ICs), the material properties, and possibly other system parameters should be established to solve the forward problem. Contrarily, inverse problems can be classified into two categories:

\begin{enumerate}[(i)]
\item Deriving the system parameters from the observed causes and effects.

\item Deriving the causes from the observed effects.
\end{enumerate}

The first category is a traditional definition of an inverse problem \citep{Inverse_problem_theory}; whereas, fundamentally, the second category is an inverse reconstruction of the full solutions using the available sparse data within a domain, i.e. measurements, with system parameters assumed to be known. The physical experiments represent the most common source of the sparse data observed within a domain. The main focus of this paper is on an inverse thermal field reconstruction, which is a task belonging to the second category. Such reconstruction is an important and often an essential part of digital twinning.

Consequently, the focus of this paper is twofold:

\begin{enumerate}[a.]

\item Developing a framework for inversely reconstructing the full temperature field from the sparse experimental measurements with speed as close to real time as possible.

\item Understanding how to generate multiple viable options satisfying given requirements, which could be used for semi-autonomous system control with a human-in-the-loop (HuIL) element \cite{huil1}.

\end{enumerate}
These two points are discussed in some detail below.

From a historical perspective inverse problems focused largely on parameter estimation for differential equations \cite{Cambridge, inverse_param, inverse_param2}, with some of the conventional methods being functional analytic regularisation as well as statistical regularisation \cite{Inverse_problem_theory, Cambridge}. Possibly the most notable example of statistical regularisation is Bayesian inversion \cite{Cambridge}. Another way to solve such inverse problems is to use a search-and-optimisation-based approach, for example the Particle Swarm Optimisation (PSO) algorithm \cite{Jaluria}. More comprehensive reviews of the various methods to solve inverse problems are provided by Tamaddon-Jahromi et al. \cite{Tamaddon_Jahromi} and Arridge et al. \cite{Cambridge}. However, the aforementioned methods tend to quickly become computationally intractable and generally lack sufficient flexibility. 

Increasingly in many engineering fields, machine learning (ML) is being applied to everything with problems ranging from manufacturing \cite{ML_manufacturing, ML_manufacturing2} to aerospace \cite{ML_aerospace, ML_aerospace2}; and, indeed, it offers a combination of unique benefits, such as accuracy, efficiency, flexibility, and scalability \cite{DL}, all of which are desirable in many engineering applications. Nevertheless, in the context of inverse analysis and digital twinning, ML models tend to have a few drawbacks.

Data-driven ML models usually require a significant amount of training data in order to provide sufficient accuracy and wide enough applicability. Although some ML model types, such as Gaussian Process Regression (GPR) \cite{GPR}, generally require far less training data than other ML models such as Neural Networks (NNs) \cite{NNs, ANNs}, Long Short-Term Memory (LSTM) \cite{LSTM}, and Transformers \cite{bielajewa2023, classic_transformer}, still a significant amount of data is needed. Collecting significant amounts of data poses a considerable challenge for complex engineering applications. This issue can be resolved in two ways: (1) Use the experimental data for the training or (2) Run a high number of standard forward simulations, FEM simulations for instance, to generate the necessary training data \cite{Tamaddon_Jahromi, bielajewa2023, GPR_thesis, ZHANG2023106354, ZHU2023120697}. The former is rarely possible in engineering, as the experimental data is usually sparse and there is not enough to be able to effectively train a data-driven ML model. The latter seems to be more promising on the surface; however, it assumes the possession of a completely accurate, verified, and trustworthy FEM model, which is also a complicated problem of its own to develop and verify, especially when dealing with extreme environments. Additionally, training ML model on selected cases of the forward simulations could bias the ML model to predict certain solutions, which might not necessarily match the physical experiment.

On the other hand, there are physics-based ML models, such as  Physics-Informed Neural Networks (PINNs), which do not rely on training data, as they are trained using relevant partial differential equations (PDEs) \cite{PINNs, PINNs_P}. This resolves the issue of potential solution bias which data-driven models are susceptible to, but creates another issue. One of the primary reasons why ML is such an attractive option for digital twinning is its high inference speed after the model is trained, which could potentially allow it to function in real time. However, PINNs require the continuous training to be performed as the sensor measurements are inputted, which slows the model down considerably. Furthermore, the integration of such physics-based ML models within existing industry simulation workflows, which are typically developed using the conventional simulation software, is not a straightforward process.

It is generally accepted that it is very challenging to directly apply FEM to inverse problems, where some or all BCs are unknown, which traditionally limit its application to forward problems only. However, the developed FEM-based inverse analysis framework demonstrate that it is possible to modify the FE workflow in such a way as to allow it solve an inverse problem and accurately reconstruct a solution.

The secondary focus of the present work is to generate multiple viable solutions in a controlled manner. The challenge of inverse problems is that they are ill-posed, meaning that the solution is non-unique \cite{Inverse_problem_theory}. However, this difficulty can be harnessed and transformed into an advantage, as this ill-posedness is surprisingly desirable when it comes to the second objective of achieving control: it can be exploited to generate multiple options (solutions) for the user.

The remainder of the paper is organised into following sections. Section~\ref{methods} details the method derivations for linear transient problem; moreover, it describes the extension of the method allowing the generation of multiple solution options fitting certain total energy criterion. Section~\ref{res_disc} presents and interprets the results for several thermal field reconstruction cases; furthermore, several solution options are produced and analysed in order to demonstrate the application of the multiple solution generation process. Finally, Section~\ref{concs} summarises the significance of the results in the context of the objectives set in Section~\ref{intro}, and also provides the main advantages of the presented framework.

\section{Methodology}\label{methods}

\begin{figure}[h]%
	\centering
	\includegraphics[width=1.0\textwidth]{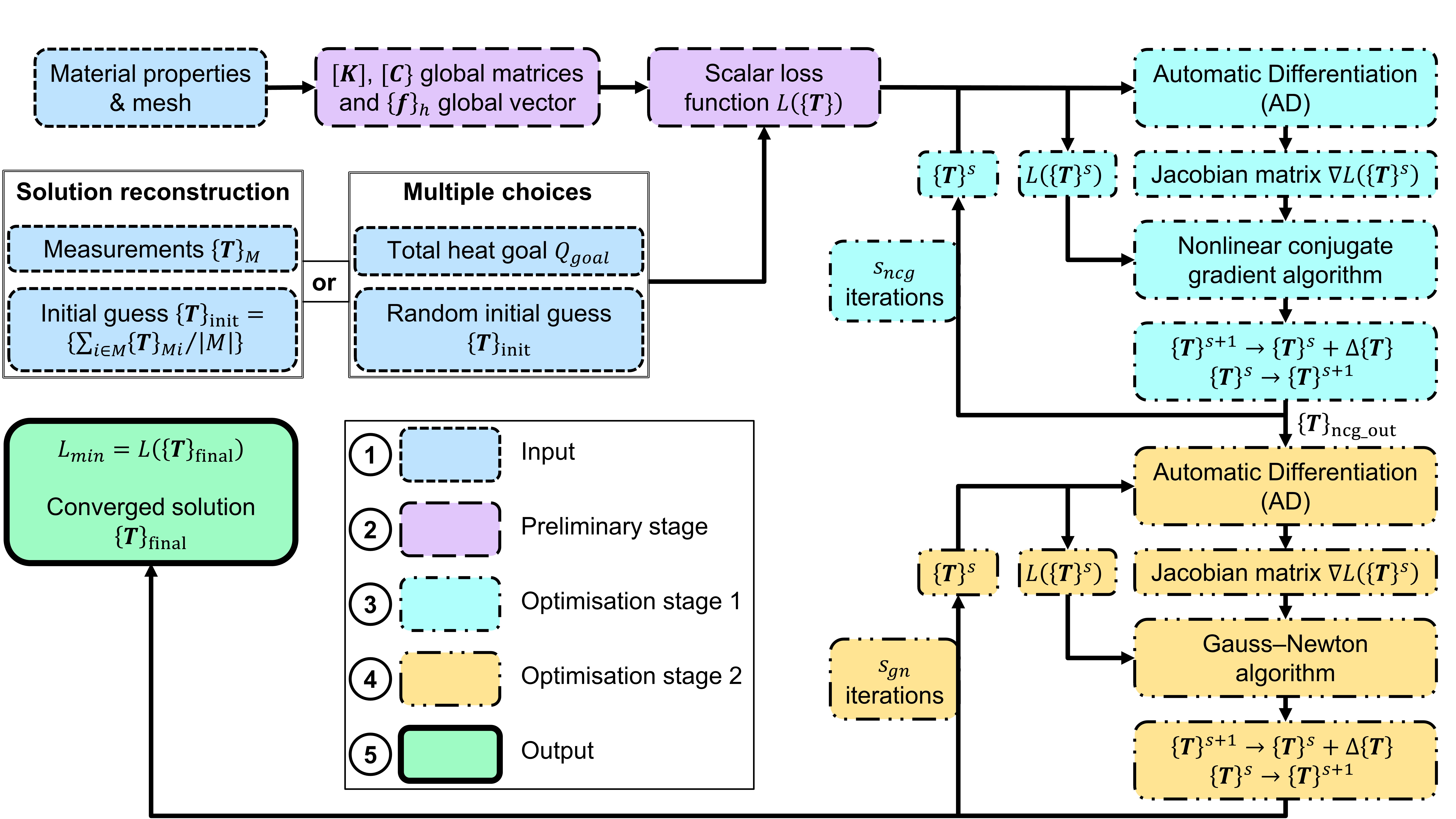}
	\caption{Summary of the workflow for the solution reconstruction and multiple solution generation. For the steady-state problem only $\left[ \bm{K} \right]$ global matrix and $\left\{ \bm{f} \right\}_{h}$ global vector are computed, and the whole procedure is repeated only once; whereas for the transient problem, it is repeated every time step and then $\left\{ \bm{T} \right\}_{\mathrm{final}}$ refers to the solution at one time step. For solution reconstruction the initial temperature distribution $\left\{ \bm{T} \right\}_{\mathrm{init}}$ is a uniform temperature equal to the average measurement value for steady state; $\left\{ \bm{T} \right\}_{\mathrm{init}}$ used for each time step is a uniform temperature equal to the average time-step measurement value.}\label{workflow_flowchart}
\end{figure}

This section provides a description of the methodology. The framework introduced in the present work is based upon the ODIL (Optimizing a DIscrete Loss) approach, which was successfully applied to fluid flow problems \cite{ODIL} and glioma radiotherapy planning \cite{ODIL_med} and has undergone substantial adaptations for this paper. Additionally, the framework is adapted to ensure compatibility with the measurements from the physical experiments. Figure~\ref{workflow_flowchart} outlines a general workflow with the details provided in the relevant sub-sections.

\subsection{Linear transient problem}\label{transient}

The general linear 3D transient heat conduction equation without a heat source and with isotropic material properties is given as \cite{FEM_Nithiarasu}:

\begin{align}
\frac{\partial}{\partial x} \left( k \frac{\partial T}{\partial x} \right) + \frac{\partial}{\partial y} \left( k \frac{\partial T}{\partial y} \right) + \frac{\partial}{\partial z} \left( k \frac{\partial T}{\partial z} \right) = \rho c_p \frac{\partial T}{\partial t}
\label{eq1}
\end{align}
where $x$, $y$, and $z$ are the spatial coordinates, $T$ is the temperature, $k$ is the thermal conductivity, while $\rho$ and $c_p$ are density and specific heat, respectively, which are assumed to be temperature-independent; $t$ is time.

\nomenclature{$x$, $y$, $z$}{Spatial coordinates}
\nomenclature{$T$}{Temperature}
\nomenclature{$k_x$, $k_y$, $k_z$}{Thermal conductivities in $x$, $y$, and $z$ directions}
\nomenclature{$k$}{Thermal conductivity}
\nomenclature{$\rho$}{Density}
\nomenclature{$G$}{Heat source}
\nomenclature{$t$}{Time}

Additionally, general expressions describing Neumann BCs are:
\begin{align}
k \frac{\partial T}{\partial x} l + k \frac{\partial T}{\partial y} m + k \frac{\partial T}{\partial z} n + q = 0 \quad \text{on} \quad \Gamma_q \nonumber \\
k \frac{\partial T}{\partial x} l + k \frac{\partial T}{\partial y} m + k \frac{\partial T}{\partial z} n + h(T-T_a) = 0 \quad \text{on} \quad \Gamma_h
\label{eq2}
\end{align}
where $l$, $m$, and $n$ are the direction cosines of the surface normals, $q$ is the time-dependent heat flux, while $h$ is a convection heat transfer coefficient, $T_a$ is ambient temperature, and $\Gamma_q$ and $\Gamma_h$ are the boundary surfaces where these BCs are applied. Dirichlet BCs are omitted in the present paper, since they are not physically representative of the experimental setting. Nonetheless, this approach can be used in conjunction with Dirichlet BCs.

\nomenclature{$l$, $m$, $n$}{Direction cosines of the surface normals}
\nomenclature{$q$}{Heat flux}
\nomenclature{$h$}{Convection heat transfer coefficient}
\nomenclature{$T_a$}{Ambient temperature}
\nomenclature{$\Gamma$}{Boundary surface}
\nomenclature{$\Gamma_q$}{Applied heat flux boundary surface}
\nomenclature{$\Gamma_h$}{Convection boundary surface}
\nomenclature{$\Omega$}{Domain}
\nomenclature{$N_i$}{Shape (interpolation) functions}

The standard Galerkin weighted residual method is employed to discretise Eq.~\ref{eq1} and Eq.~\ref{eq2} in space \cite{FEM_Nithiarasu}; and fully implicit time discretisation scheme is used. Consequently, the following global system of equations is obtained:
\begin{align}
	\left[ \bm{C} \right] \left\{ \frac{\left\{ \bm{T} \right\}^{n+1} - \left\{ \bm{T} \right\}^n}{\Delta t} \right\} + \left[ \bm{K} \right]  \left\{ \bm{T} \right\}^{n+1}  = \left\{ \bm{f} \right\}^{n+1}
	\label{eq16}
\end{align}
with
\begin{align}
\left[ \bm{K} \right] = \int_\Omega \left[ \bm{B} \right]^T
\left[ \bm{D} \right] \left[ \bm{B} \right] \mathrm{d} \Omega + \int_{\Gamma_h} h \left[ \bm{N} \right]^T \left[ \bm{N} \right] \mathrm{d} \Gamma
\label{eq6}
\end{align}
\begin{align}
\left[ \bm{C} \right]= \int_\Omega \rho c_p \left[ \bm{N} \right]^T \left[ \bm{N} \right] \mathrm{d} \Omega
\label{eq13}
\end{align}
\begin{align}
\left\{ \bm{f} \right\} = - \int_{\Gamma_q} q \left[ \bm{N} \right]^T \mathrm{d} \Gamma + \int_{\Gamma_h} h T_a \left[ \bm{N} \right]^T \mathrm{d} \Gamma
\label{eq7}
\end{align}
where $\left\{ \bm{T} \right\}$ and $\left\{ \bm{N} \right\}$ are the temperature and shape function vectors, respectively.  $\left[ \bm{C} \right]$ and $\left[ \bm{K} \right]$ are the global mass and stiffness matrices, respectively, and $\left\{ \bm{f} \right\}$ is a global loading vector. Superscript $n$ signifies the $n^\text{th}$ time step; $\Omega$ is the discretised domain.

\nomenclature{$\left[ \bm{C} \right]$}{Global mass matrix}
\nomenclature{$\left[ \bm{K} \right]$}{Global stiffness matrix}
\nomenclature{$\left\{ \bm{f} \right\}$}{Global loading vector}
\nomenclature{$\theta$}{Time discretisation parameter}
\nomenclature{$\left\{ \bm{f} \right\}_h$}{Global convection loading vector}
\nomenclature{$\left\{ \bm{f} \right\}_q$}{Global applied heat flux loading vector}

The loading term $\left\{ \bm{f} \right\}$ can be split into two following terms corresponding to the known (convection) and unknown (applied heat flux) BCs - global convection $\left\{ \bm{f} \right\}_h$ and applied heat flux loading vectors $\left\{ \bm{f} \right\}_q$, respectively:
\begin{align}
\left\{ \bm{f} \right\} = \left\{ \bm{f} \right\}_q + \left\{ \bm{f} \right\}_h \nonumber \\
\left\{ \bm{f} \right\}_q = - \int_{\Gamma_q} q \left[ \bm{N} \right]^T \mathrm{d} \Gamma \nonumber \\
\left\{ \bm{f} \right\}_h = \int_{\Gamma_h} h T_a \left[ \bm{N} \right]^T \mathrm{d} \Gamma
\label{eq10}
\end{align}
Eq~\ref{eq10} represents a start of the novel deviation of the developed inverse analysis approach from the standard FEM workflow used for forward problems. The global applied heat flux vector $\left\{ \bm{f} \right\}_q$ is assumed be unknown.

\subsection{Inverse analysis}\label{inverse_analysis}

The following assumptions are made for the inverse analysis:
\begin{itemize}
\item The temperature values are known at points belonging to the measurement set $M \in \Omega$.
\item The materials properties are known; they include the general thermal conductivity matrix $\left[ \bm{D} \right]$ and the convection heat transfer coefficient $h$.
\item The ambient temperature $T_a$ is known.
\item The exact locations of $\Gamma_h$ and $\Gamma_q$ on the boundary are known, and $\Gamma = \Gamma_h \cup \Gamma_q$, where $\Gamma$ is domain ($\Omega$) boundary.
\end{itemize}

Given that $M$ number of temperatures $\left\{ \bm{T} \right\}_M $ are known from measurements and the heat flux value on the boundary $\Gamma_q$ is unknown, the next step is to define a scalar loss function $L$ of $\left\{ \bm{T} \right\}^{n+1}$. The dependency of $\left\{ \bm{f} \right\}_{q}$ on the temperature distribution $\left\{ \bm{T} \right\}$ is as follows (Eqs.~\ref{eq16} and~\ref{eq10}): 
\begin{align}
\left\{ \bm{f} \right\}^{n+1}_q = \left[ \bm{C} \right] \left\{ \frac{\left\{ \bm{T} \right\}^{n+1} - \left\{ \bm{T} \right\}^n}{\Delta t} \right\} + \left[ \bm{K} \right]  \left\{ \bm{T} \right\}^{n+1}  - \left\{ \bm{f} \right\}^{n+1}_{h}
\label{eq17}
\end{align}
Then $L \left( \left\{ \bm{T}^{n+1} \right\} \right)$ is a sum of three terms:
\begin{enumerate}[(a)]
    \item Residual term $ \\ c_1 \sum_{i \in \Omega \setminus \Gamma_q} \left( \left[ \bm{C} \right]_i \left\{ \frac{\left\{ \bm{T} \right\}^{n+1} - \left\{ \bm{T} \right\}^n}{\Delta t} \right\} + \left[ \bm{K} \right]_i  \left\{ \bm{T} \right\}^{n+1}  - \left\{ \bm{f} \right\}^{n+1}_{hi} \right)^2 \\$
    where $\left[ \bm{C} \right]_i$ and $\left[ \bm{K} \right]_i$ are the $i^\text{th}$ rows of the global $\left[ \bm{C} \right]$ and $\left[ \bm{K} \right]$ matrices, respectively, which correspond to the $i^\text{th}$ node in the domain; while $\left\{ \bm{f} \right\}_{hi}$ is the $i^\text{th}$ element in the global $\left\{ \bm{f} \right\}_{h}$ vector. This term enforces the fact that, in the absence of the volumetric heating, the elements of loading vector $\left\{ \bm{f} \right\}_{q}$ corresponding to the nodes inside the domain and the nodes where only the convection is applied should be zero (see~\ref{secA1}).
    \item Measurement term $ c_2 \sum_{i \in M} \left( \left\{ \bm{T} \right\}^{n+1}_i - \left\{ \bm{T} \right\}^{n+1}_{Mi}\right)^2 \\$
    This term directly incorporates the temperature measurements into the simulation. 
    \item Regularisation (smoothing) term 
    $\\ c_3 \left[ \sum_{i \in \Gamma_q \setminus \Gamma_{edge}} \left( \left\{ \bm{f} \right\}^{n+1}_{qi} - \frac{\sum_{i \in \Gamma_q \setminus \Gamma_{edge} }  \left\{ \bm{f} \right\}^{n+1}_{qi} }{\left|\Gamma_q \setminus \Gamma_{edge} \right|} \right)^2 + \right. \\ 
    \\ + \sum_{i \in \Gamma_{edge} \setminus \Gamma_{corners}} \left( \left\{ \bm{f} \right\}^{n+1}_{qi} - \frac{\sum_{i \in \Gamma_{edge} \setminus \Gamma_{corners} } \left\{ \bm{f} \right\}^{n+1}_{qi} }{\left|\Gamma_{edge} \setminus \Gamma_{corners} \right|} \right)^2 +\\
    \\ \left. + \sum_{i \in \Gamma_{corners}} \left( \left\{ \bm{f} \right\}^{n+1}_{qi} - \frac{\sum_{i \in \Gamma_{corners} } \left\{ \bm{f} \right\}^{n+1}_{qi} }{\left|\Gamma_{corners} \right|} \right)^2 \right]\\$
    $\left\{ \bm{f} \right\}^{n+1}_{q}$ is calculated using Eq.~\ref{eq17}. $\Gamma_{edge}$ represents the 3D edge of the surface where the heat flux $q$ is applied, i.e. it is $\partial \Gamma_{q}$ and also, referring to the last point in the aforementioned assumptions list, it is equal to $\Gamma_q \cap \Gamma_h$. $\Gamma_{corners}$ represents sharp features on $\Gamma_{edge}$ if such features exist and if they are discretised using one element, i.e the node on a corner belongs only to one element. An example of $\Gamma_{edge}$ and $\Gamma_{corners}$ is given in Table~\ref{table00}. This term ensures that there is a smooth temperature transition on the boundary of the domain where the heat flux $q$ is applied.
\end{enumerate}

In the above equations, $c_1$, $c_2$, and $c_3$ are weighting coefficients signifying the relative importance of each term; the selection of the appropriate values is discussed in Section~\ref{res_disc}. Mesh, as well as $\left[ \bm{C} \right]$, $\left[ \bm{K} \right]$, and $\left\{ \bm{f} \right\}_h$ can be obtained using virtually any commercial or open-source FEM software, such as ANSYS \cite{ansys} or Code\_Aster \cite{code_aster}; the latter is used in this paper. The component in the form of $\left[ \bm{C} \right] \left\{ \left( \left\{ \bm{T} \right\}^{n+1} - \left\{ \bm{T} \right\}^n \right) / \Delta t \right\}$, which signifies heat accumulation over time, becomes zero for steady-state problems.

\nomenclature{$L$}{Scalar loss function}
\nomenclature{$c_1, c_2, c_3, c_4$}{Loss function weighting coefficients}
\nomenclature{$\Gamma_{edge}$}{3D edge of $\Gamma_q$ surface}
\nomenclature{$\Gamma_{corners}$}{Sharp features on $\Gamma_{edge}$ discretised using one element}

The regularisation (smoothing) term prevents unphysically large and abrupt temperature variations on the surface from appearing. It does so through indirectly ensuring the smooth spatial variation of temperature's first derivative normal to the surface. The first derivative of temperature in the direction normal to the boundary directly correlates with the heat flux through the boundary $q = -k (\partial T/ \partial n )$ \cite{FEM_Nithiarasu}, which in turn defines the applied heat flux loading vector $\left\{ \bm{f} \right\}_{q}$. The regularisation term limits the deviation of each element of $\left\{ \bm{f} \right\}_{q}$ from the mean value of all $\left\{ \bm{f} \right\}_{q}$ elements. It is necessary to apply the regularisation term only to $\Gamma_q$, where the implicit degrees of freedom in the form of the unknown applied heat flux are situated.

The regularisation term is split into three parts: the first one corresponding to the nodes on $\Gamma_q \setminus \Gamma_{edge}$, the second one corresponding to $\Gamma_{edge} \setminus \Gamma_{corners}$, and the third one to $\Gamma_{corners}$. When the applied heat flux varies gradually on $\Gamma_q$ relative to the finite element size, a rapid change in the loading vector $\left\{ \bm{f} \right\}_{q}$ values usually occurs between the nodes belonging to $\Gamma_q \setminus \Gamma_{edge}$, the nodes belonging to $\Gamma_{edge} \setminus \Gamma_{corners}$ and $\Gamma_{corners}$ if such nodes exist. Therefore, in order to avoid the complications related to this sudden transition, it is beneficial to separate the regularisation term into aforementioned groups. Four-node quadrilateral (linear) elements on $\Gamma_q$ are used in conjunction with the regularisation term defined above; however, a slightly different node grouping within this loss function term might be necessary if three-point (linear) triangles, particularly unstructured ones, are to be used for $\Gamma_q$ surface discretisation. Additionally, certain adjustments would need to be made for the meshes with mesh refinement on $\Gamma_q$ in order to account for the potential presence of varying element size.

Lastly, it should be noted that such regularisation term definition might not necessarily be an optimal choice for all potential heat flux distributions. However, considering a specific experimental arrangement, reasonable assumptions usually can be made regarding a general shape of the heat flux distribution produced by a given heat-generating element; meaning that some knowledge can be derived regarding the relative magnitudes of heat fluxes without knowing their specific values. For example, if the heat flux is produced by an induction coil of known geometry located at a certain position relative to a test piece, then some plausible assumptions can be made regarding the approximate locations of regions of the maximum and minimum heat fluxes. This information can subsequently be used to customise the regularisation term.

In order to select the correct distribution out of the infinite number of possible distributions fitting the residual term, the measurement and regularisation terms are employed in conjunction, thus making the inverse problem approximately well-posed. The heat flux associated with a particular temperature distribution is implicitly calculated within the regularisation term using the node equations excluded from the residual term (part (a) of the loss function and~\ref{secA1}) to obtain the loading vector values corresponding to the nodes on $\Gamma_q$.

While the current paper utilises only linear analysis, the extension of this framework to nonlinear material properties naturally follows. It is slightly simplified by the fact that unlike stress-strain curve, thermal material properties, $k$, $\rho c_p$, and $h$, tend to steadily increase with temperature without displaying a softening region. Therefore, the potential additional solution non-uniqueness stemming purely from the nonlinear material properties \cite{FEM_Zienkiewicz_V2} might be avoided.

\subsection{Minimisation}\label{minimisation}

After the loss function $L \left(  \left\{ \bm{T} \right\}^{n+1} \right)$ is defined, the main objective becomes the minimisation of $L$, and $\left\{ \bm{T} \right\}$ corresponding to the minimum of $L$ would be the solution for the $n+1$ time step, i.e. the reconstructed temperature field at this time step. It should be emphasised that the minimum of $L$ might not be necessarily zero, as, depending on its definition, the regularisation term might not be required to be exactly zero for the correct solution. The function minimisation is a common optimisation problem, but it is a cornerstone in the ML training process, and hence it is advantageous to borrow some of the methods frequently used in ML to minimise $L$.

Figure~\ref{workflow_flowchart} summarises a general workflow for the solution reconstruction for each time step. In this paper two iterative gradient-based optimisation algorithms are used consecutively: the first one is nonlinear conjugate gradient method (Stage 1) for $s_{ncg}$ iterations \cite{CG} and the second one is Gauss–Newton method (Stage 2) for $s_{gn}$ iterations \cite{GN1, GN2}. The output from the nonlinear conjugate gradient method $\left\{ \bm{T} \right\}_{\text{ncg\_out}}$ is an input to Gauss–Newton method. The nonlinear conjugate gradient method utilises a first-order derivative matrix (Jacobian matrix) $\nabla L \left( \left\{ \bm{T} \right\} \right)$, whilst the Gauss–Newton method makes use of the approximation of the second-order derivative matrix (Hessian matrix) $\nabla^2 L \left( \left\{ \bm{T} \right\} \right)$ based on $\nabla L \left( \left\{ \bm{T} \right\} \right)$. Similar to the ML training process, Automatic Differentiation (AD) is employed in order to efficiently calculate the exact Jacobian matrix at each iteration \cite{ODIL, AD}. The initial temperature distributions used for each time step is chosen to be a uniform temperature equal to the average time-step measurement value. At the beginning of each iteration $ L \left( \left\{ \bm{T} \right\}^s \right)$ and $\nabla L \left( \left\{ \bm{T} \right\}^s \right)$ are calculated for the temperature distribution outputted by the previous iteration $\left\{ \bm{T} \right\}^s$, then $\Delta \left\{ \bm{T} \right\}$ is computed using either nonlinear conjugate gradient method (during Stage 1) or Gauss–Newton method (during Stage 2). Finally, the temperature distribution for the next iteration becomes $ \left\{ \bm{T} \right\}^{s+1} = \left\{ \bm{T} \right\}^s + \Delta \left\{ \bm{T} \right\}$.

An iterative minimisation of the residual term combined with the measured temperature values and the regularisation term will provide a temperature distribution over the entire domain including boundaries. The temperature distribution along the heat flux boundary $\Gamma_q$ can now be used to compute the heat flux.

To ensure a high computational speed the majority of matrix additions, multiplications, as well as AD are performed on Graphical Processing Unit (GPU) with the help of PyTorch, which is a highly optimised ML library \cite{PyTorch}.

\nomenclature{$s_{ncg}$}{No. of iterations for nonlinear conjugate gradient method}
\nomenclature{$s_{gn}$}{No. of iterations for Gauss–Newton method}

\subsection{Multiple choice generation}\label{mult_choices}

In the previous sub-sections, the regularisation term is used to overcome the inherent ill-posedness of the inverse problem ensuring that the loss function converges to the correct solution; whereas this section demonstrates how ill-posedness can be exploited in order to generate multiple solutions fitting certain predefined requirements. A scenario can be considered where some desired values of the time-dependent total heat $Q_{\mathrm{goal}} (t)$ should be achieved, which can be done by applying the heat flux on $\Gamma_q$ in various ways. It is assumed that the measurements are not available, as the aim of this procedure is to produce potential future options which can be used for control, rather than to reconstruct the correct solution from the obtained measurements.

\nomenclature{$Q_{\mathrm{goal}}$}{Desired values of total heat}

For the transient problem, the loss function is defined as a sum of the following three terms, two of which have been used previously for the solution reconstruction:

\begin{enumerate}
    \item Residuals term $ \\ c_1 \sum_{i \in \Omega \setminus \Gamma_q} \left( \left[ \bm{C} \right]_i \left\{ \frac{\left\{ \bm{T} \right\}^{n+1} - \left\{ \bm{T} \right\}^n}{\Delta t} \right\} + \left[ \bm{K} \right]_i  \left\{ \bm{T} \right\}^{n+1}  - \left\{ \bm{f} \right\}^{n+1}_{hi} \right)^2 $.
    \item Regularisation (smoothing) term 
    $\\ c_3 \left[ \sum_{i \in \Gamma_q \setminus \Gamma_{edge}} \left( \left\{ \bm{f} \right\}^{n+1}_{qi} - \frac{\sum_{i \in \Gamma_q \setminus \Gamma_{edge} }  \left\{ \bm{f} \right\}^{n+1}_{qi} }{\left|\Gamma_q \setminus \Gamma_{edge} \right|} \right)^2 + \right. \\ 
    \\ + \sum_{i \in \Gamma_{edge} \setminus \Gamma_{corners}} \left( \left\{ \bm{f} \right\}^{n+1}_{qi} - \frac{\sum_{i \in \Gamma_{edge} \setminus \Gamma_{corners} } \left\{ \bm{f} \right\}^{n+1}_{qi} }{\left|\Gamma_{edge} \setminus \Gamma_{corners} \right|} \right)^2 +\\
    \\ + \sum_{i \in \Gamma_{corners}} \left( \left\{ \bm{f} \right\}^{n+1}_{qi} - \frac{\sum_{i \in \Gamma_{corners} } \left\{ \bm{f} \right\}^{n+1}_{qi} }{\left|\Gamma_{corners} \right|} \right)^2 +\\
    + \left. \left( \frac{\sum_{i \in \Gamma_{edge} \setminus \Gamma_{corners} } \left\{ \bm{f} \right\}^{n+1}_{qi} }{\left|\Gamma_{edge} \setminus \Gamma_{corners} \right|} - 2 \cdot \frac{\sum_{i \in \Gamma_{corners} } \left\{ \bm{f} \right\}^{n+1}_{qi} }{\left| \Gamma_{corners} \right|} \right) \right]. \\$
    The dependency of $\left\{ \bm{f} \right\}^{n+1}_{q}$ on the temperature distribution $\left\{ \bm{T} \right\}$  is is given by Eq.~\ref{eq17}.
    \item Total heat term $ \\ c_4  \left( \sum_{i \in \Gamma} \left( \left[ \bm{C} \right]_i \left\{ \frac{\left\{ \bm{T} \right\}^{n+1} - \left\{ \bm{T} \right\}^n}{\Delta t} \right\} + \left[ \bm{K} \right]_{\mathrm{base}\_i}  \left\{ \bm{T} \right\}^{n+1}  \right) - Q_{\mathrm{goal}}^{n+1}  \right)^2$
\end{enumerate}
where
\begin{align}
\left[ \bm{K} \right]_{\mathrm{base}} = \int_\Omega \left[ \bm{B} \right]^T
\left[ \bm{D} \right] \left[ \bm{B} \right] \mathrm{d} \Omega
\label{eq22}
\end{align}
and $\left[ \bm{K} \right]_{\mathrm{base}\_i}$ is the $i^\text{th}$ row of the $\left[ \bm{K} \right]_{\mathrm{base}}$ matrix.

The total heat term serves to enforce the required total heat values for the solution as it progresses through time. $c_4$ represents the relative importance of the total heat specification; a lower $c_4$ value means the generated solution is allowed to deviate from the desired total heat values more, and vice versa. It can be used to control the type of the solution generated. The randomisation of the initial temperature values allows the generation of different solutions for the same $c_4$ value, which is explained in greater detail in this section.

The regularisation term includes an additional fourth component, which limits the magnitude of the deviation between the average value of $\left\{ \bm{f} \right\}_q$ elements on $\Gamma_{edge} \setminus \Gamma_{corners}$ and the average value of $\left\{ \bm{f} \right\}_q$ elements on $\Gamma_{corners}$. The separation of the regularisation term into three parts, which is beneficial for the solution reconstruction, becomes less effective when generating multiple solutions, as there is no measurement term creating a connection between the three regions. Therefore, if the fourth component is absent from the regularisation term, the generated solutions tend to exhibit sharp temperature variations between the corners and the rest of $\Gamma_q$. The fourth component serves to constrain the applied heat flux on the corners without overly constraining the problem overall, so that it becomes well-posed.

Figure~\ref{workflow_flowchart} summarises a general workflow for generating multiple solutions, here the total heat term replaces the measurement term used for the solution reconstruction. A similar approach of combining two gradient-based optimisation algorithms, conjugate gradient and Gauss–Newton (Sub-section~\ref{minimisation}), is adopted for the loss function minimisation. The only difference with the solution reconstruction process is the initial temperature distributions used for each time step. In theory, any gradient-free optimisation algorithm can be used as a wrapper for the two aforementioned gradient-based algorithms in order to find multiple local minima of the loss function landscape, and basin-hopping \cite{basinhopping} was initially trialled for this purpose. Nevertheless, just randomised assignment of the temperature values at each node works well while keeping the adherence to the runtime specification for the digital twinning. The random initial temperature values are produced using probability density function of the uniform distribution over the half-open interval $[T_{min}, T_{max})$ \cite{probability}. A new random set of values for the initial guess vector $\left\{ \bm{T} \right\}_{\mathrm{init}}$ is produced for every time step using the same $T_{min}$ and $T_{max}$ values for the whole run; the random $\left\{ \bm{T} \right\}_{\mathrm{init}}$ corresponding to each time step are generated prior to the run. It should be noted that $T_{min}$ and $T_{max}$ values limit only the initial guess, and they do not limit the final converged temperature distribution.

Finally, while only the total heat requirement is considered in the present work, it is easy enough to create a loss function term enforcing any other given requirement, such as the required maximum temperature on $\Gamma_q$.

\section{Results and discussion}\label{res_disc}

\begin{figure}[!b]%
\centering
\includegraphics[width=0.5\textwidth]{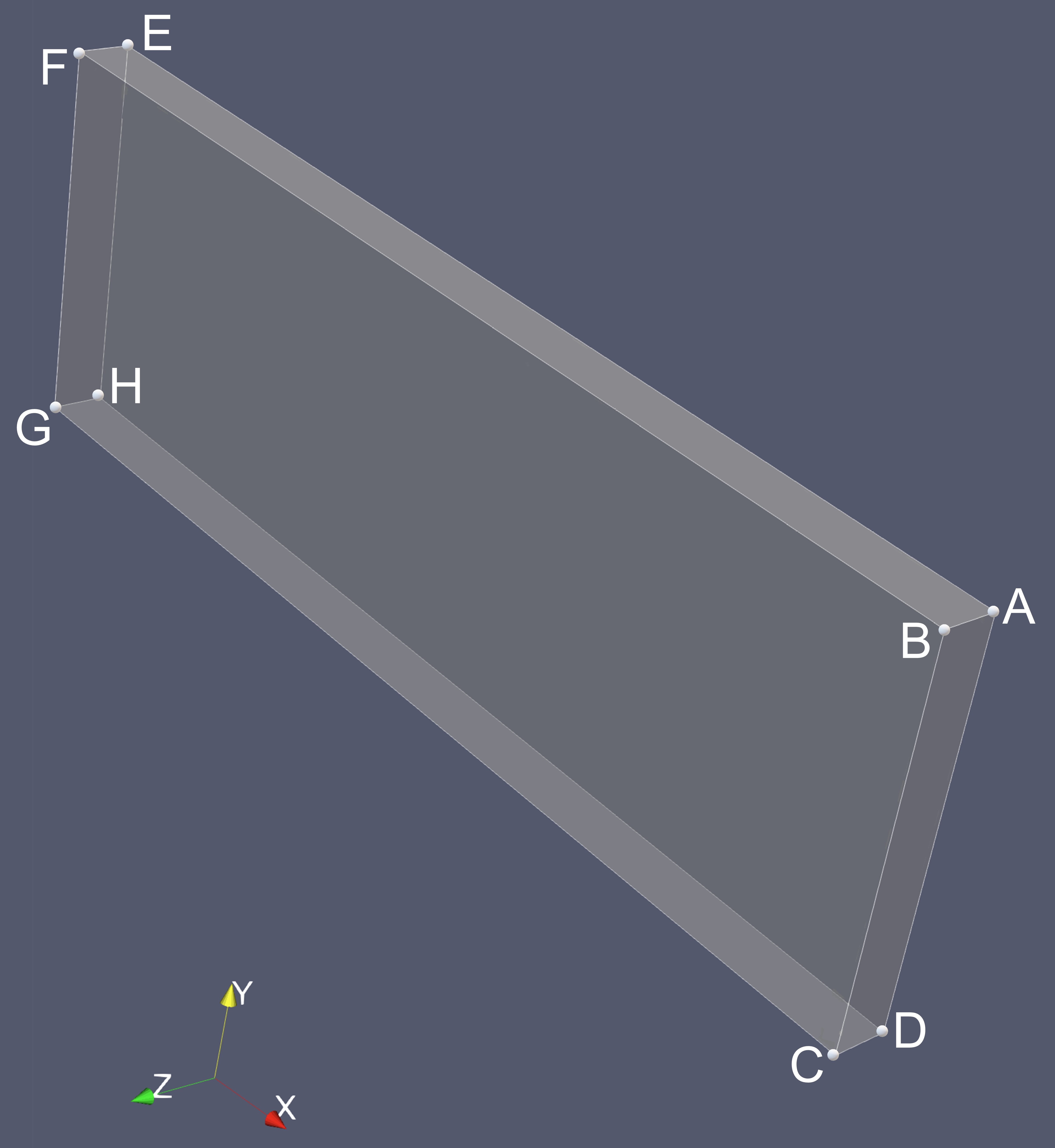}
\caption{Labels for the applied BCs (Table~\ref{table0}).}\label{plate_general}
\end{figure}

The example considered in this paper is an air-cooled stainless steel plate subjected to the linearly increasing uniform surface heating given by: 
\begin{align}
q(t) = \frac{600000}{180} t
\label{eq18}
\end{align}
Figure~\ref{plate_general} shows a general schematic of the plate with the applied BCs described in Tables~\ref{table0} and~\ref{table00}, while Table~\ref{table1} provides the material properties. This particular geometry and BC types are selected as they represent a generally realistic experimental setup. The simulation is run for 180s; the initial conditions represent a steady state, room temperature of 20$^\circ \text{C}$. For this case $\Gamma_{edge}$ is ABCD edge shown in Figure~\ref{plate_general} and also equal to $\Gamma_q \cap \Gamma_h$; $\Gamma_{corners}$ consists of the nodes A, B, C, and D (Table~\ref{table00}).

\begin{table}[!b]
\begin{threeparttable}[!b]
\begin{tabular}{p{2cm}p{3.8cm}}
  \toprule
\textbf{Surface} & \textbf{BC} \\
  \midrule
  ABCD & Uniform heat flux \\
  \midrule
  BFGC & \multirow{5}{3.8cm}{Plate-air convection} \\
  BFEA & \\
  AEHD & \\
  DHGC & \\
  EFGH & \\
   \toprule
\end{tabular}
\caption{Applied BCs; Figure~\ref{plate_general} shows the labels.}\label{table0}
\end{threeparttable}
\end{table}

\begin{table}[!b]
\begin{threeparttable}[!b]
\begin{tabular}{p{2cm}p{3.8cm}}
  \toprule
\textbf{Set} & \textbf{Location} \\
  \midrule
  $\Gamma_q$ & Surface ABCD \\
  $\Gamma_{edge}$ & Edge ABCD \\
  $\Gamma_{corners}$ & Nodes A, B, C, D\\
   \toprule
\end{tabular}
\caption{Defined sets; Figure~\ref{plate_general} shows the labels.}\label{table00}
\end{threeparttable}
\end{table}

\begin{table}[!b]
\begin{threeparttable}[!b]
\begin{tabular}{p{9cm}p{2cm}p{2cm}}
  \toprule
\textbf{Parameter} & \textbf{Value} & \textbf{Unit} \\
  \midrule
Thermal conductivity $k$ (stainless steel Grade 91) & 25.84 & $W/(m \: ^\circ \text{C})$ \\
Air-steel convection heat transfer coefficient $h$ & 135.00 & $W/(m^2 \: ^\circ \text{C})$\\
Density $\rho$ (stainless steel Grade 91) & 7760.00 & $kg/(m^3)$ \\
Specific heat $c_p$ (stainless steel Grade 91) & 416.80 & $J/(kg \: ^\circ \text{C})$ \\
Atmospheric temperature $T_a$ & 20.00 & $^\circ \text{C}$ \\
   \toprule
\end{tabular}
\caption{Material properties.}\label{table1}
\begin{tablenotes}
\item[*] All values are given for the room temperature of 20$^\circ \text{C}$.
\end{tablenotes}
\end{threeparttable}
\end{table}

\subsection{Thermal field reconstruction}\label{ther_field_rec}

The reference solutions, to which the inversely reconstructed solutions are compared, are generated in Code\_Aster \cite{code_aster} by applying the BCs shown in Table~\ref{table0}. In order to inversely reconstruct the solution from the sparse measurements, the steady-state loss function is firstly defined (Sub-section~\ref{inverse_analysis}) and used to reconstruct the steady-state solution, which is utilised as a starting point for the subsequent transient calculations. Then, the transient loss function can be defined (Sub-section~\ref{inverse_analysis}) and used to reconstruct the solution at each time step.

Table~\ref{table2} gives the descriptions of the six cases considered in the present sub-section. $\Delta t_{\mathrm{ref}}$ and $\Delta t_{\mathrm{rec}}$ are time step sizes used for generating the reference solution in Code\_Aster and for the solution reconstruction process, respectively.

\nomenclature{$\Delta t_{\mathrm{ref}}$}{Time step size used for generating reference solution}
\nomenclature{$\Delta t_{\mathrm{rec}}$}{Time step size used for solution reconstruction process} 

Parameters $c_1$, $c_2$ are equal to 1.0 for the solution reconstruction; conversely, $c_3$ might need a separate adjustment, the process of which is detailed in~\ref{secA2}. The locations of $c_3$ value regions resulting in the lowest relative errors seem to be dependent on the number of measurements as well as on $\Delta t_{\mathrm{rec}}$; consequently, it can be calibrated prior to the experiment. Moreover, the only meaningful values $c_3$ might adopt are between 0.0 and 1.0, as any value above 1.0 means that $c_1$ and $c_2$ can be adjusted in order to keep $c_1$, $c_2$, and $c_3$ below 1.0. Generally, values of $c_3$ above 0.5 seem to yield acceptable errors. 

\begin{figure}[ht]
    \centering
    \begin{minipage}[b]{0.48\linewidth}
        \includegraphics[width=\linewidth]{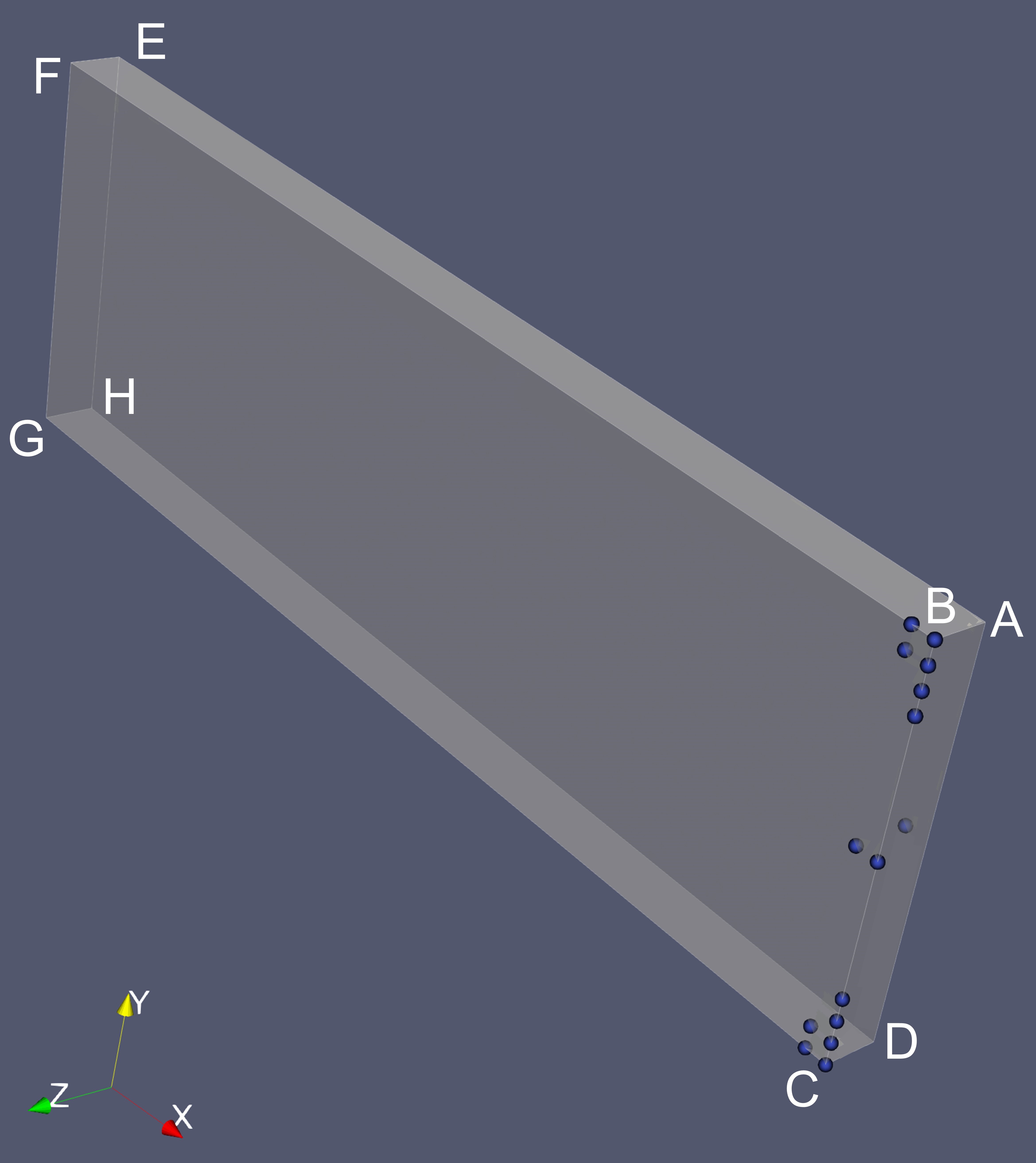}
    \end{minipage}
    \hspace{0.3cm} % Space between images
    \begin{minipage}[b]{0.48\linewidth}
        \includegraphics[width=\linewidth]{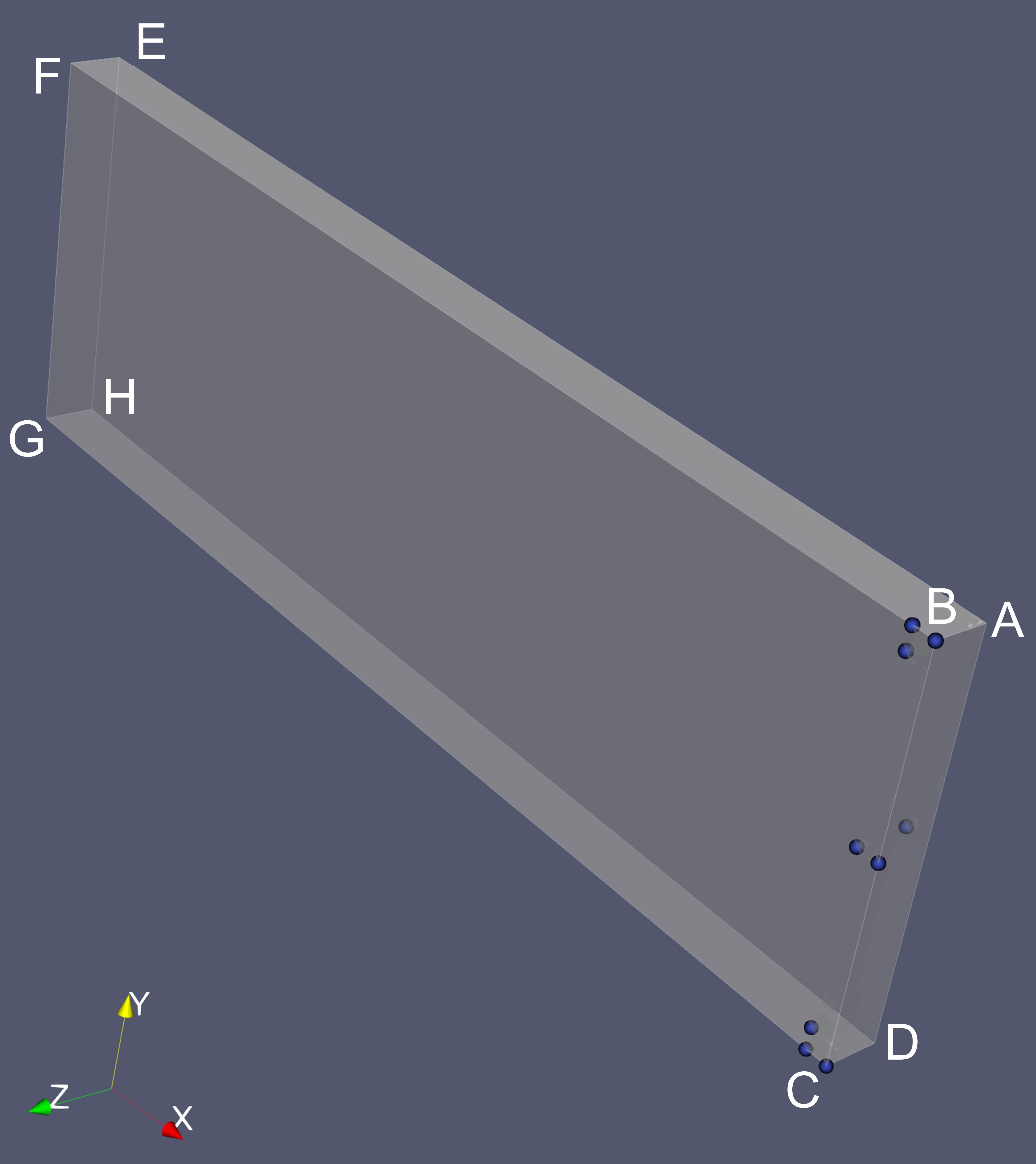}
    \end{minipage}
    \caption{Two options for the measurement placement considered: 15 (left) and 9 (right) measurements.}\label{meas}
\end{figure}

One structured mesh consisting of 5,196 linear quadrilateral and hexahedral elements (for surface and volume discretisation, respectively) is used for the reference solutions as well as for all six cases of the solution reconstruction. During the calculation of elemental matrices and vectors, four Gauss (integration) points are used for the surface integration of quadrilaterals, and eight Gauss points are used for the volume integration of hexahedrals. No mesh refinement is used in the present paper. Two options for the measurement placement are considered: 15 and 9 measurements; Figure~\ref{meas} shows the locations for each one. Furthermore, these measurements are experimentally attainable when thermocouple (TC) measurements and possibly infrared (IR) camera surface recordings are available \cite{Hancock2018, Chimera2}. None of the measurements are located directly on $\Gamma_q \setminus \Gamma_{edge}$, as it is generally not possible to place any diagnostic sensors directly under the heat-generating element, for example an induction coil \cite{Hancock2018}. Furthermore, none of the measurements are placed close to EFGH side (Figure~\ref{plate_general}), since, with the location of the heating element being known, this region would not provide any useful information for a significant part of the simulation in the beginning as the temperatures measured there would simply remain nearly constant.

%1. Case16 (Case 11) - temperature gradient regularisation, matching delta_t with the ref sol 1s, more mes points
%2. Case 13 - temperature gradient regularisation, matching delta_t with the ref sol 1s, less mes points
%3. Case 14 - like Case 16 only ref solution generated using 0.1s time step
%4. Case 15 - like Case 13 only ref solution generated using 0.1s time step
%5. Case 17 - like Case 14 only the reconstructed soluton time step is 2s
%6. Case 18 - like Case 15 only the reconstructed soluton time step is 2s

\begin{table}[!b]
\small
\begin{threeparttable}[!b]
\caption{Description of six cases considered for the stainless steel plate.}
\begin{tabular}{p{0.8cm}|p{0.9cm}|p{0.9cm}p{2.0cm}p{0.8cm}p{0.8cm}p{0.8cm}p{0.8cm}p{0.8cm}}
  \toprule
Case No. & $\Delta t_{\mathrm{ref}}$ [s] \tnote{a} & $\Delta t_{\mathrm{rec}}$ [s] \tnote{b} & No. of measurements & $s_{ncg}$ \tnote{c} & $s_{gn}$ \tnote{d} & $c_1$ & $c_2$ & $c_3$\\
  \midrule
1 & 1.0 & 1.0 & 15 & 2 & 1 & 1.0 & 1.0 & 1.0 \\
2 & 1.0 & 1.0 & 9 & 2 & 1 & 1.0 & 1.0 & 1.0 \\
  \midrule
3 & 0.1 & 1.0 & 15 & 2 & 1 & 1.0 & 1.0 & 1.0 \\
4 & 0.1 & 1.0 & 9 & 2 & 1 & 1.0 & 1.0 & 1.0 \\
  \midrule
5 & 0.1 & 2.0 & 15 & 2 & 1 & 1.0 & 1.0 & 1.0 \\
6 & 0.1 & 2.0 & 9 & 2 & 1 & 1.0 & 1.0 & 0.5 \\
   \toprule
\end{tabular}
\label{table2}
\begin{tablenotes}
\item[a] $\Delta t_{\mathrm{ref}}$ is a time step size used for generating the reference solution in Code\_Aster. 
\item[b] $\Delta t_{\mathrm{rec}}$ is a time step size used for the solution reconstruction process.
\item[c] $s_{ncg}$ is a number of nonlinear conjugate gradient algorithm iterations (Figure~\ref{workflow_flowchart}).
\item[d] $s_{gn}$ is a number of Gauss–Newton algorithm iterations (Figure~\ref{workflow_flowchart}).
\end{tablenotes}
\end{threeparttable}
\end{table} 

\begin{figure}[!b]
\centering
\includegraphics[width=1.0\textwidth]{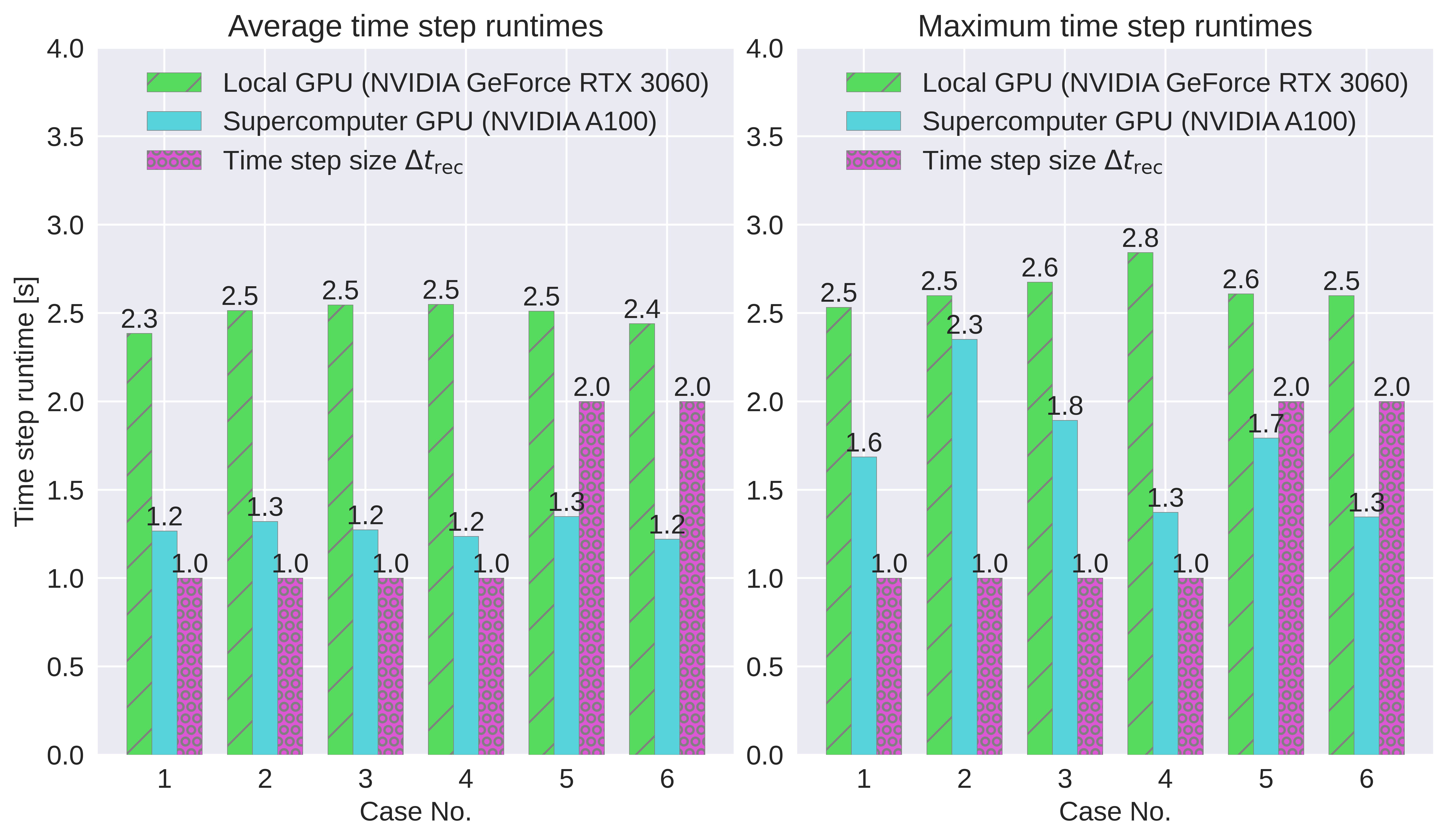}
\caption{Average and maximum time step runtimes for all cases, compared with the time step size $\Delta t_{\mathrm{rec}}$ used; one GPU is used for the local and supercomputer calculations.}\label{runtimes_all_cases}
\end{figure}

\begin{figure}[!b]
\centering
\includegraphics[width=1.0\textwidth]{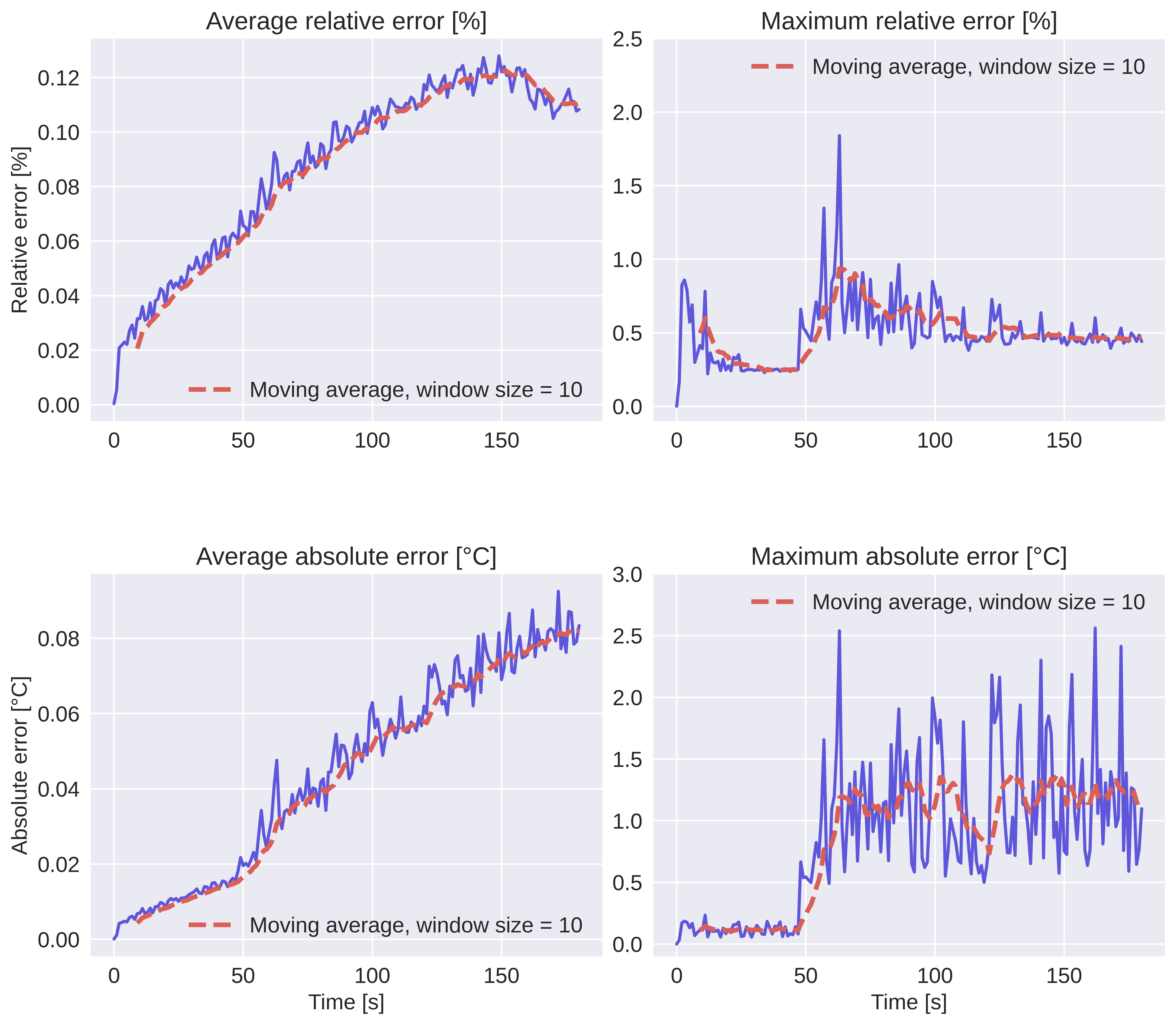}
\caption{The dependence of relative and absolute errors on time for Case No. 1.}\label{case1_errors}
\end{figure}

\begin{figure}[!b]
\centering
\includegraphics[width=1.0\textwidth]{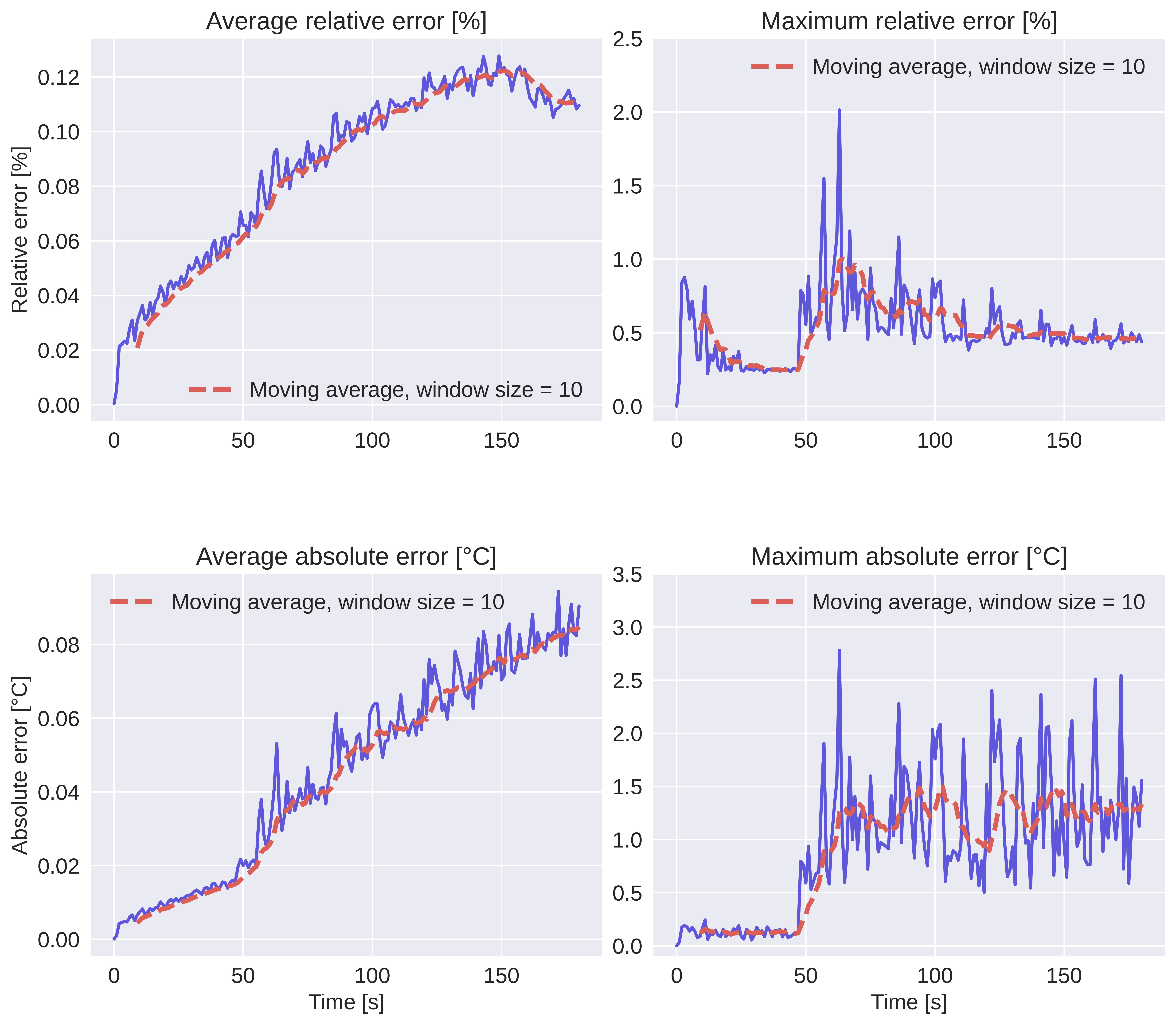}
\caption{The dependence of relative and absolute errors on time for Case No. 2.}\label{case2_errors}
\end{figure}

The accuracy is judged by calculating the errors of the reconstructed solution relative to the reference solution, while the computational speed is evaluated by recording the time it takes to reconstruct the temperature field at each time step, i.e. the time step runtime. Table~\ref{table3} summarises the overall relative and absolute errors averaged in space and time as well as the overall maximum relative and absolute errors for each case, whereas Figure~\ref{runtimes_all_cases} compares the average time step runtime for local GPU (NVIDIA GeForce RTX 3060) and supercomputer GPU (NVIDIA A100) with the time step size $\Delta t_{\mathrm{rec}}$ for each case. Table~\ref{table3} also provides relative and absolute differences between two reference solutions used in this paper, the first one with $\Delta t_{\mathrm{ref}}$ equal to 1s and the second one with $\Delta t_{\mathrm{ref}}$ equal to 0.1s. Moreover, Figures~\ref{case1_errors} and \ref{case2_errors} provide a more detailed view of the relative and absolute errors as they develop with time for Cases No. 1 and 2, respectively; the errors on these figures are averaged in space only. Similar figures for Cases No. 3-6 can be found in~\ref{secA3}. Since all calculations are performed on the same mesh, the space averaging is simply done using the corresponding nodal temperature values.

\begin{table}[!b]
\small
\begin{threeparttable}[!b]
\caption{Average and maximum relative and absolute solution reconstruction errors for six cases considered for the stainless steel plate.}
\begin{tabular}{p{0.7cm}|p{2.5cm}p{2.5cm}|p{2.9cm}p{2.9cm}}
  \toprule
Case No. & Average relative error [\%] & Maximum relative error [\%] & Average absolute error [$^\circ \text{C}$] & Maximum absolute error [$^\circ \text{C}$] \\
  \midrule
1 & 0.09 & 1.84 & 0.05 & 2.56 \\
2 & 0.09 & 2.01 & 0.05 & 2.78 \\
  \midrule
3 & 0.14 & 1.77 & 0.06 & 2.54 \\
4 & 0.14 & 1.98 & 0.06 & 2.83 \\
  \midrule
5 & 0.21 & 1.75 & 0.08 & 1.88 \\
6 & 0.21 & 1.67 & 0.09 & 2.58 \\
   \midrule
Ref. diff. \tnote{a} & 0.15 & 1.33 & 0.07 & 1.0 \\
   \toprule
\end{tabular}
\label{table3}
\begin{tablenotes}
\item[a] Reference difference row - this row corresponds to the relative and absolute difference between the two reference solutions, the first one with $\Delta t_{\mathrm{ref}}$ equal to 1s and the second one with $\Delta t_{\mathrm{ref}}$ equal to 0.1s. 
\end{tablenotes}
\end{threeparttable}
\end{table}

\subsubsection{Number of measurements and errors}\label{no_meas-errors}

Firstly, the correlation between the number of measurements and the error distribution is considered. The six cases can be divided into three pairs, in which all parameters are the same apart from the number of measurements, excluding $c_3$ in some cases as it is calibrated based on the number of measurements and $\Delta t_{\mathrm{rec}}$ (\ref{secA2}): Cases No. 1 and 2, Cases No. 3 and 4, and Cases No. 5 and 6 (Table~\ref{table2}). Unsurprisingly, the overall maximum relative errors increase with the decrease in the number of measurements for Cases No. 1 and 2, Cases No. 3 and 4 (Table~\ref{table3}); on the other hand, for Cases No. 5 and 6 the maximum relative error decreased by 0.08\% with the decrease in the number of measurements.

\begin{figure}[!b]
\centering
\includegraphics[width=1.0\textwidth]{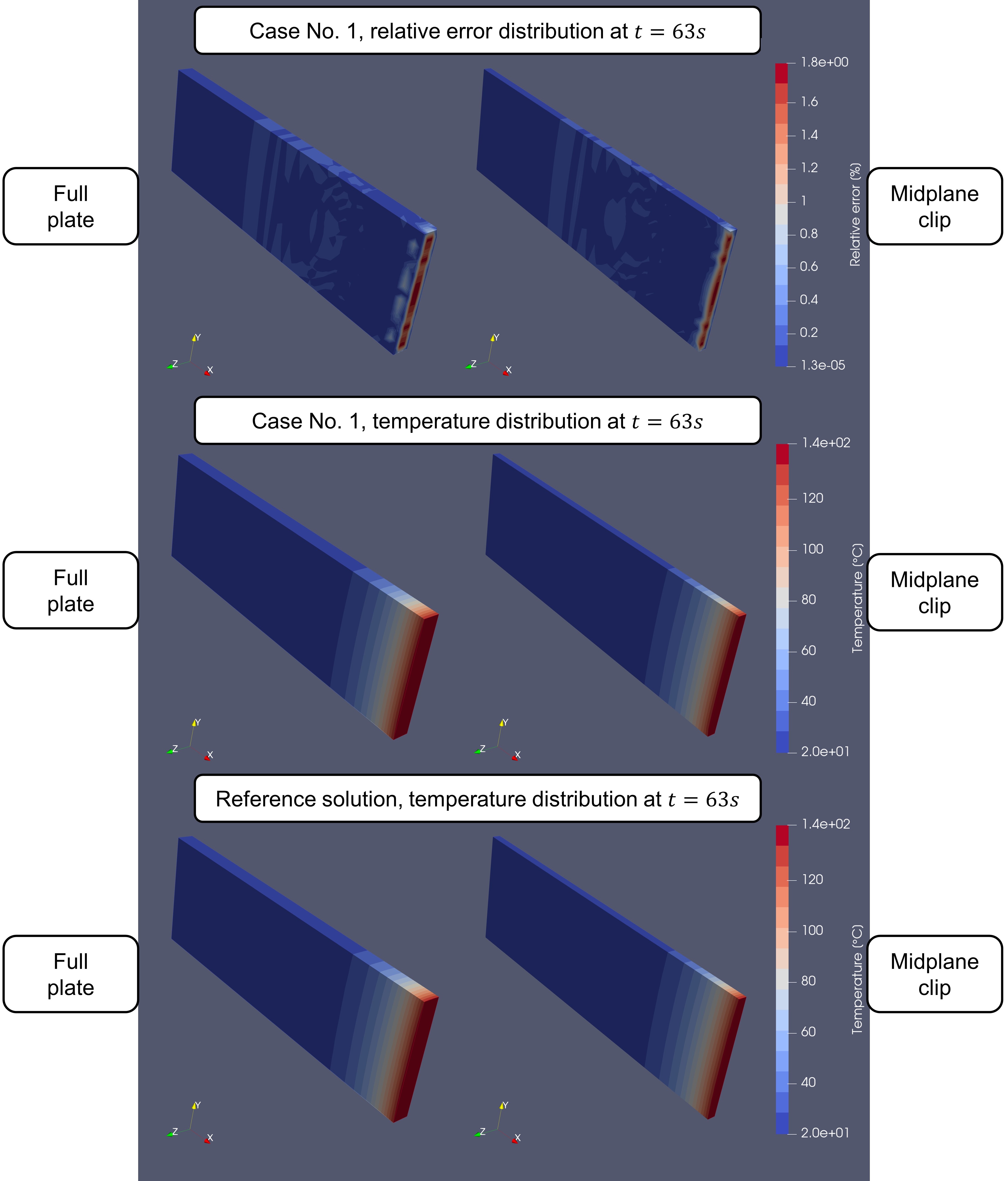}
\caption{The relative error and temperature distributions for the 63s time instance of Case 1, together with the reference solution temperature distribution; this time instance corresponds to the maximum overall relative error for Case 1.}\label{errors_case1_timestep}
\end{figure}

\begin{figure}[!b]
\centering
\includegraphics[width=1.0\textwidth]{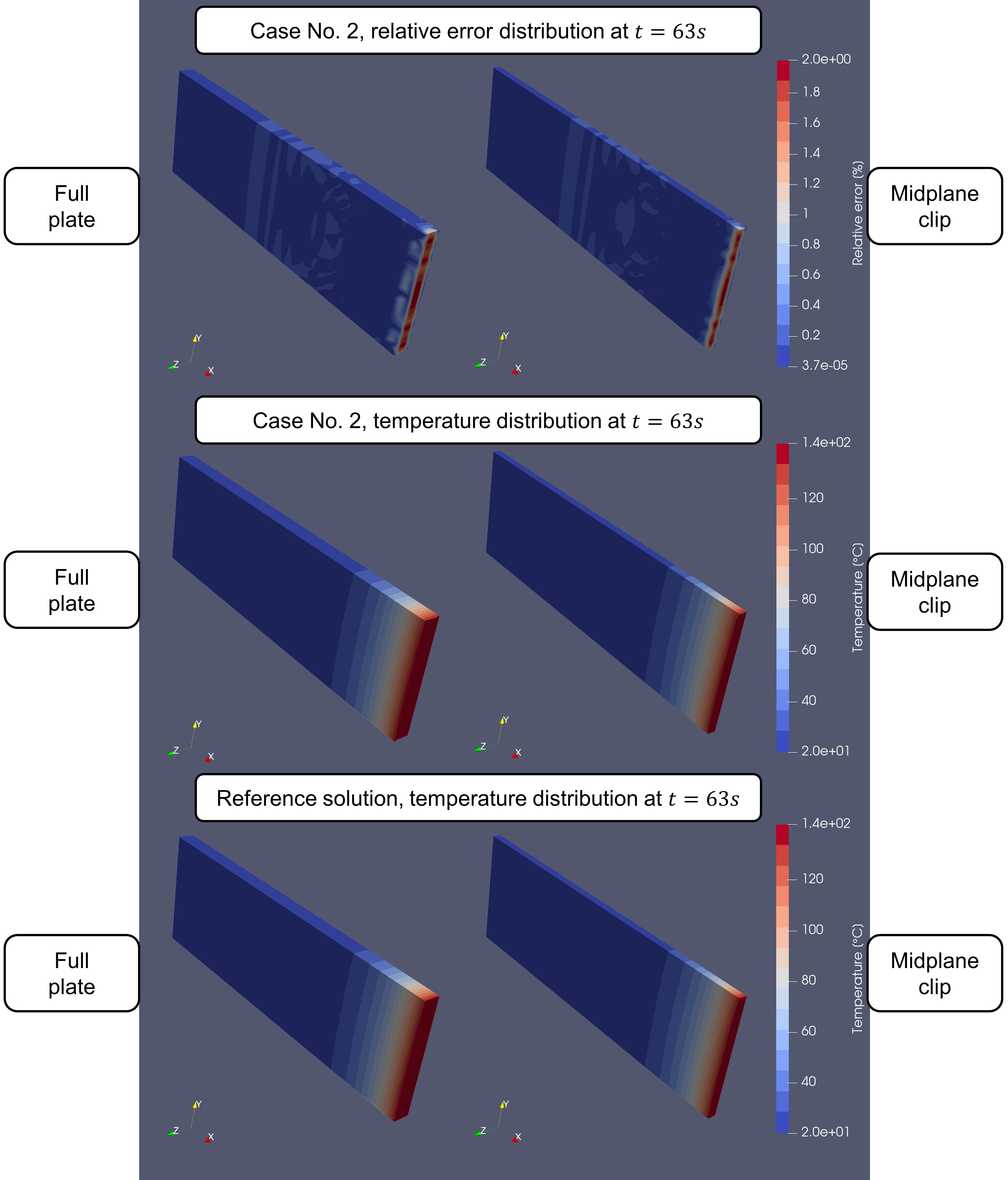}
\caption{The relative error and temperature distributions for the 63s time instance of Case 2, together with the reference solution temperature distribution; this time instance corresponds to the the maximum overall relative error for Case 2.}\label{errors_case2_timestep}
\end{figure}

Figures~\ref{errors_case1_timestep} and~\ref{errors_case2_timestep} display the spatial relative error distributions for the time instance with the maximum overall relative error for Cases No. 1 and 2, respectively; the time instance is 63s for both of them. It can be seen that the areas with the highest errors are around $\Gamma_q \setminus \Gamma_{edge} $ (Figure~\ref{plate_general}, Table~\ref{table00}), which is to be expected as no measurements are available there and no relationship between $T$ and $q$ is known a priori. However, it can be observed that the errors are still within the 2\% range, and Figures~\ref{errors_case1_timestep} and~\ref{errors_case2_timestep} show an excellent match between the reconstructed and reference solutions.

As shown in Figures~\ref{case1_errors} and~\ref{case2_errors}, as well as in Figures~\ref{case3_errors} and~\ref{case4_errors}, and Figures~\ref{case5_errors} and~\ref{case6_errors} in~\ref{secA3}, the average and maximum relative errors do not increase continuously with time, but in fact peak at the time instance between 0 and 180s and then start to gradually decline. This signifies that the measurements do provide a limiting effect on the error accumulation in time.

\subsubsection{$\Delta t_{\mathrm{ref}}$, $\Delta t_{\mathrm{rec}}$, and errors}\label{delta_t-errors}

The second set of interesting results is related to the time step ratio (TSR), which can be defined as follows:
\begin{align}
\mathrm{TSR} = \frac{\Delta t_{\mathrm{rec}}}{\Delta t_{\mathrm{ref}}}
\label{eq20}
\end{align}

\begin{figure}[!b]
\centering
\includegraphics[width=1.0\textwidth]{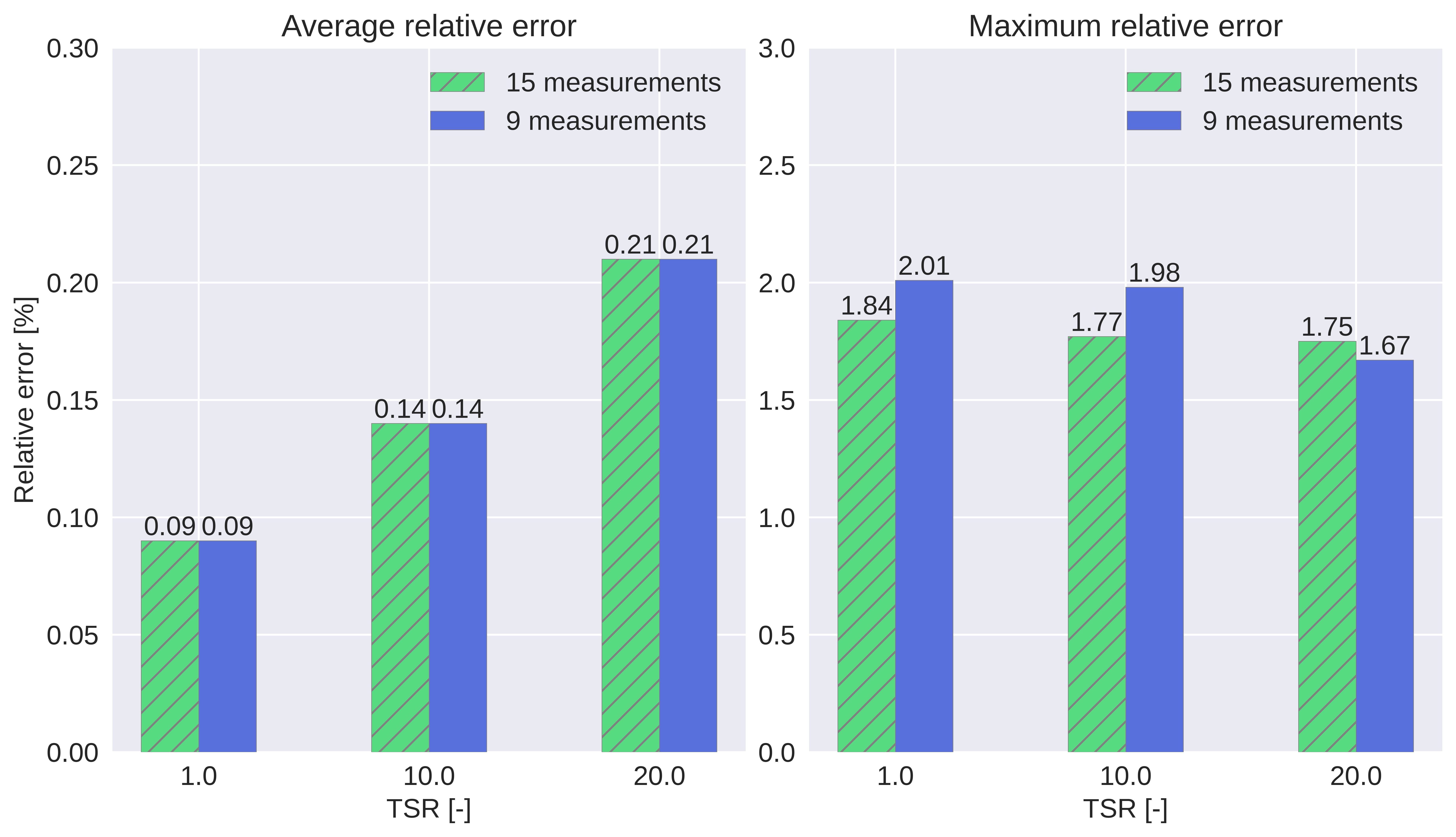}
\caption{The dependence of the average and maximum relative errors on TSR; 15 measurements are used in Cases 1, 3, and 5 (1, 10, and 20 TSR, respectively), while 9 measurements are used in Cases 2, 4, and 6 (1, 10, and 20 TSR, respectively). }\label{TSR_error}
\end{figure}

For Cases No. 1 and 2 $\mathrm{TSR}$ is equal to 1.0, for Cases No. 3 and 4 it is equal to 10, while for Cases No. 5 and 6 to 20; six cases can be divided into two groups, in which all parameters are the same apart from the TSR: Cases No. 1, 3, 5 with 15 measurements, and Cases No. 2, 4, 6 with 9 measurements (Table~\ref{table2}). Figure~\ref{TSR_error} visualises the results from Table~\ref{table3} for these two groups. It is evident that the average and maximum errors increase with a much lower rate than the corresponding TSR, especially considering the relative difference between two reference solutions used in this paper, the first one with $\Delta t_{\mathrm{ref}}$ equal to 1s and the second one with $\Delta t_{\mathrm{ref}}$ equal to 0.1s. There is 0.15\% average and 1.33\% maximum relative difference between two reference solutions (Table~\ref{table3}); however, it is apparent from Figure~\ref{TSR_error} that errors do not increase by nearly these amounts as TSR progresses from 1 to 20, indeed the maximum relative error remain nearly constant. A tenfold increase in TSR resulted in 1.6 times increase in relative average error; however, the maximum relative errors decrease with increasing TSR for both sets of measurements. This observation is highly beneficial for two reasons:

\begin{enumerate}
    \item It is yet another proof that the measurements incorporated into the algorithm provide a limiting effect on the solution reconstruction errors.
    \item The time step size can be increased to decrease the runtime, to meet real time requirements, since it can be reasonably increased with only minor effect on the reconstruction.
\end{enumerate}

\subsubsection{Runtimes}\label{runtimes}

Figure~\ref{runtimes_all_cases} compares the average and maximum time step runtimes with the time step sizes $\Delta t_{\mathrm{rec}}$ used for the solution reconstruction. For the digital twinning application it is essential that the calculation speed is as close as possible to real time; thus, the time step runtime should be equivalent to or less than the time step size. While the runtimes are significantly shorter than the values typically found in literature for the physics-based methods \cite{PINNs_P, ODIL}, the utilisation of the local GPU is still not quite enough to match the time step size. On the other hand, the supercomputer GPU runtimes are in the range of 1.2-1.8s with $\Delta t_{\mathrm{rec}}$ being equal to 2s (Cases No. 5 and 6); hence, the solution reconstruction on supercomputer happens in near real time with 1-2s fixed delay. 

This changes for nonlinear and more complex problems. For meshes with a higher number of nodes several GPUs can be utilised concurrently to sufficiently reduce runtimes, the process which might not be dissimilar to how ML models can be trained on multiple GPUs.

\subsection{Multiple choices}\label{mult_choices2}

In this sub-section it is assumed that the desired time-dependent total heat $Q_{\mathrm{goal}}$ is as follows:
\begin{align}
Q_{\mathrm{goal}} (t) = - 0.03 t^2 + 1.7 t
\label{eq21}
\end{align}
The transient loss function is defined as described in sub-section~\ref{mult_choices}.

Table~\ref{table4} presents the selected parameter values. It is decided to use $\Delta t_{\mathrm{rec}}$ equal to 2s for generating multiple solutions, as this value coupled with the current time step runtimes allows to emulate a near real-time operation with a constant delay as it is shown in sub-section~\ref{ther_field_rec}. Three sets of randomised initial guesses are produced corresponding to the three different generated solutions for each value of $c_4$, with each set containing the number of initial guess vectors $\left\{ \bm{T} \right\}_{\mathrm{init}}$ equal to the number of time steps. The simulation is run for 30s. 

\begin{table}[!b]
\small
\begin{threeparttable}[!b]
\caption{Parameter  values used for generating multiple solution options.}
\begin{tabular}{p{0.8cm}|p{0.9cm}p{0.9cm}p{0.8cm}p{0.8cm}p{0.8cm}p{0.8cm}p{0.8cm}p{0.9cm}p{1.2cm}}
  \toprule
Opt. No. & $\Delta t_{\mathrm{rec}}$ [s] \tnote{a} & $T_{min}$ [$^\circ \text{C}$]& $T_{max}$ [$^\circ \text{C}$]& $s_{ncg}$ \tnote{c} & $s_{gn}$ \tnote{d} & $c_1$ & $c_3$ & $c_4$ & Random set No.\\
  \midrule
1 & 2.0 & 20 & 100 & 2 & 1 & 1.0 & 0.1 & 0.11 & 1 \\
2 & 2.0 & 20 & 100 & 2 & 1 & 1.0 & 0.1 & 0.11 & 2 \\
3 & 2.0 & 20 & 100 & 2 & 1 & 1.0 & 0.1 & 0.11 & 3 \\
  \midrule
4 & 2.0 & 20 & 100 & 2 & 1 & 1.0 & 0.1 & 0.10 & 1 \\
5 & 2.0 & 20 & 100 & 2 & 1 & 1.0 & 0.1 & 0.10 & 2 \\
6 & 2.0 & 20 & 100 & 2 & 1 & 1.0 & 0.1 & 0.10 & 3 \\
  \midrule
7 & 2.0 & 20 & 100 & 2 & 1 & 1.0 & 0.1 & 0.14 & 1 \\
8 & 2.0 & 20 & 100 & 2 & 1 & 1.0 & 0.1 & 0.14 & 2 \\
9 & 2.0 & 20 & 100 & 2 & 1 & 1.0 & 0.1 & 0.14 & 3 \\
   \toprule
\end{tabular}
\label{table4}
\begin{tablenotes}
\item[a] $\Delta t_{\mathrm{rec}}$ is a time step size used for generating a solution option.
\item[c] $s_{ncg}$ is a number of nonlinear conjugate gradient algorithm iterations (Figure~\ref{workflow_flowchart}).
\item[d] $s_{gn}$ is a number of Gauss–Newton algorithm iterations (Figure~\ref{workflow_flowchart}).
\end{tablenotes}
\end{threeparttable}
\end{table} 

\subsubsection{Total heat}

\begin{figure}[!b]
\centering
\includegraphics[width=1.0\textwidth]{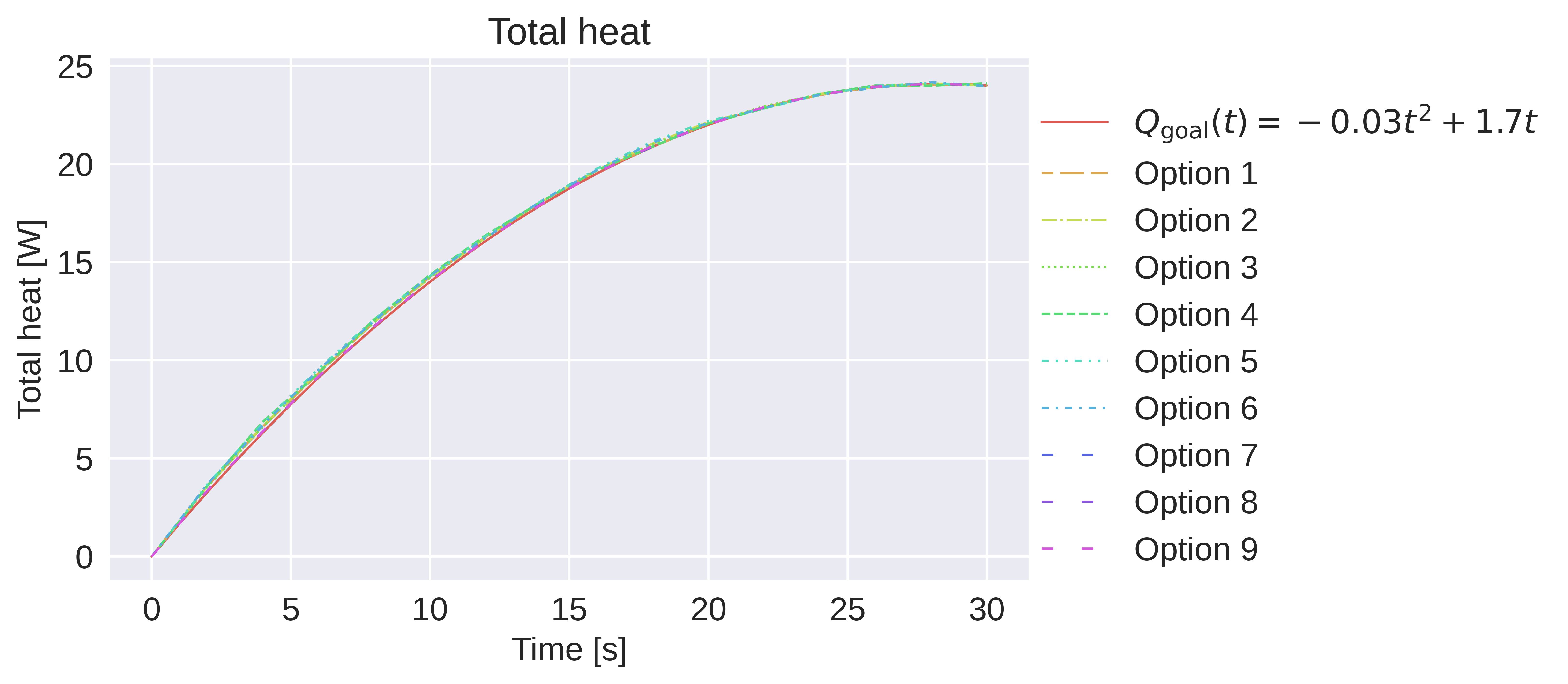}
\caption{The dependence of the total heat on time for various solution options; the total heat goal $Q_{\mathrm{goal}}$ corresponds to Eq.~\ref{eq21}.}\label{total_heat}
\end{figure}

\begin{figure}[!b]
\centering
\includegraphics[width=1.0\textwidth]{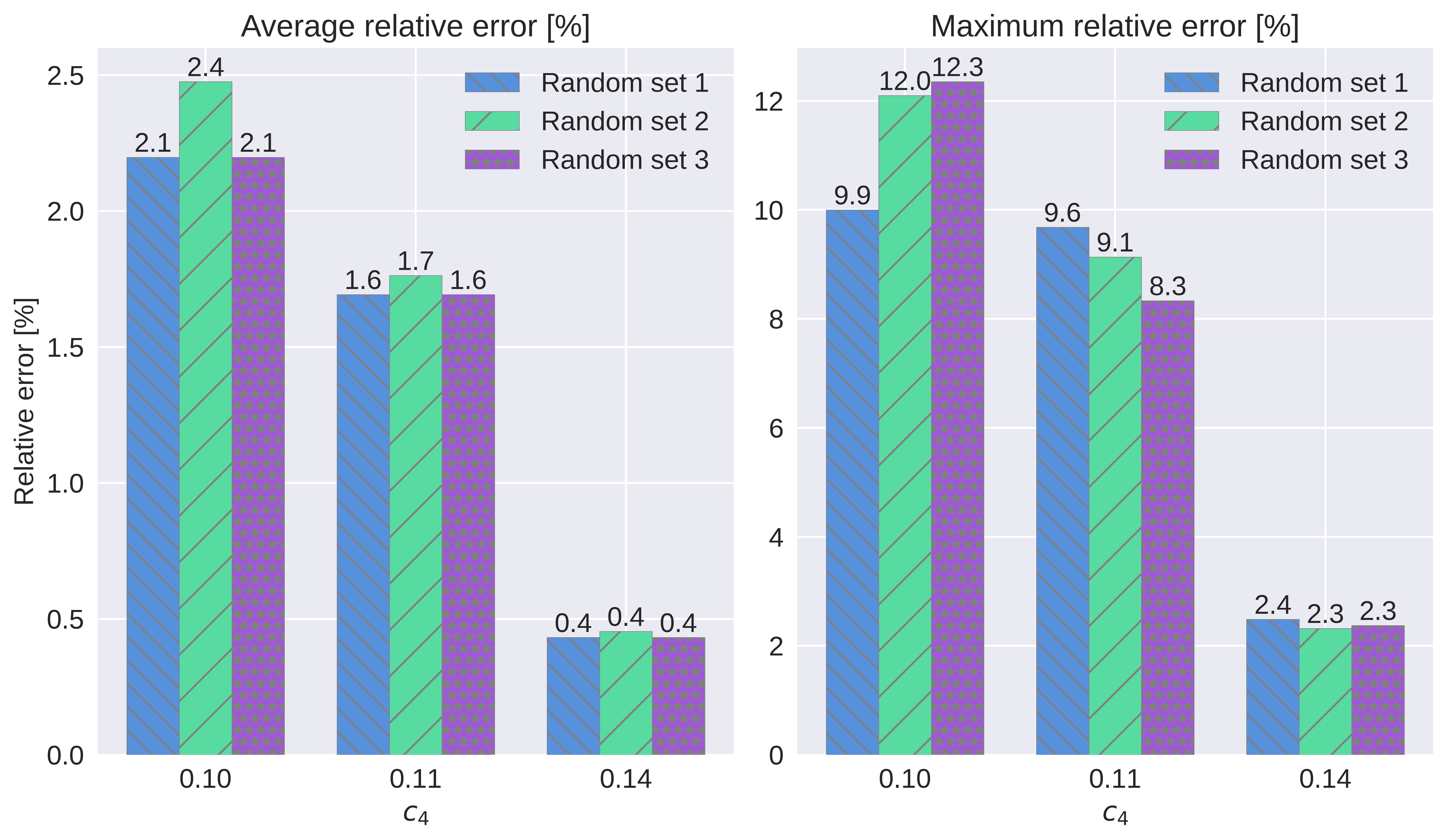}
\caption{The dependence of the relative total heat errors on $c_4$.}\label{total_heat_sets}
\end{figure}

\begin{figure}[!b]
\centering
\includegraphics[width=1.0\textwidth]{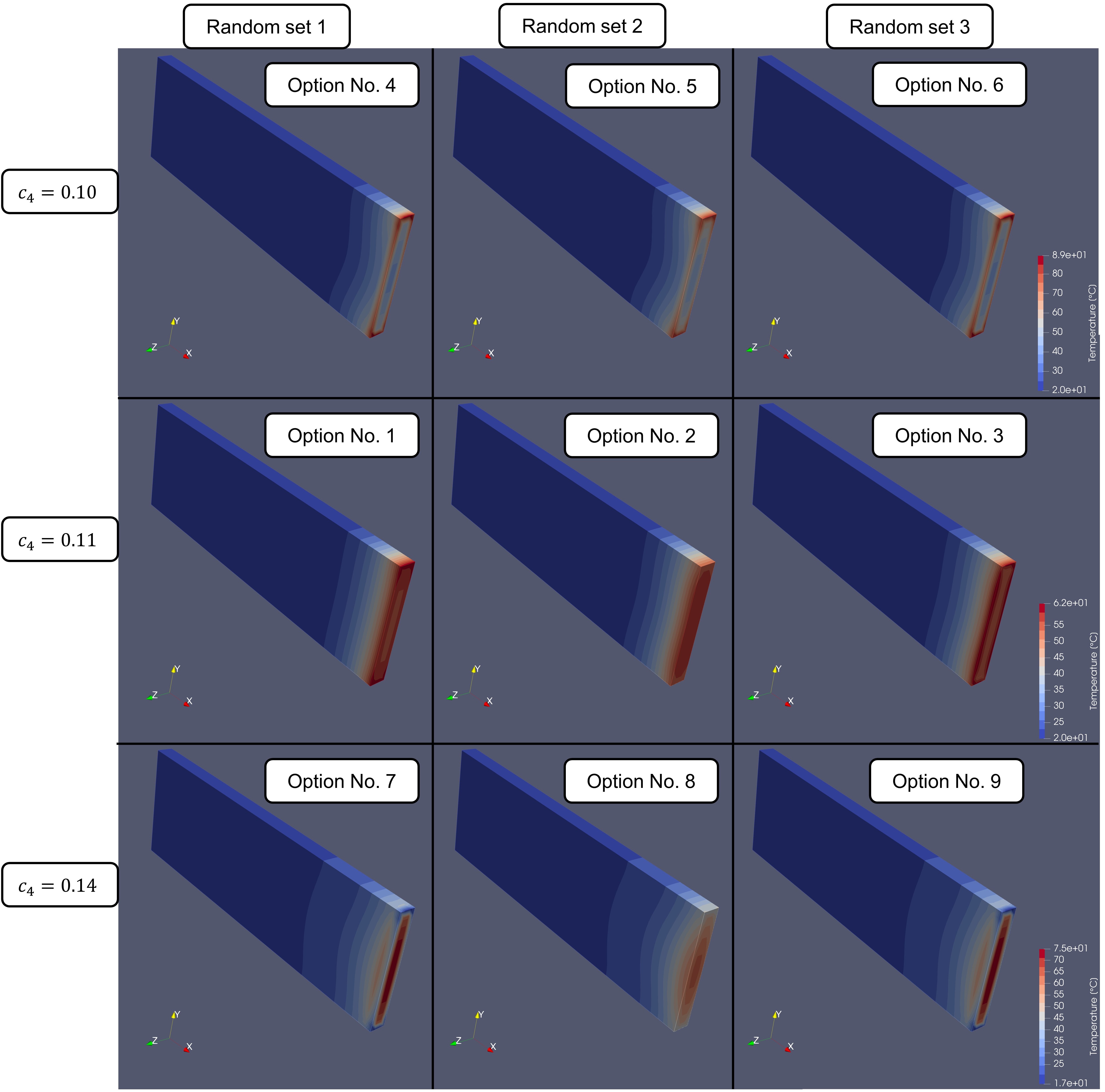}
\caption{Temperature distributions for nine solution options at 26s time instance.}\label{options_timesteps}
\end{figure}

Figure~\ref{total_heat} shows a total heat variation with time for each of the nine solution options; it is compared with the total heat goal $Q_{\mathrm{goal}} (t)$ represented by Eq.~\ref{eq21}. Table~\ref{table4_1} summarises the overall total heat relative and absolute errors averaged in time as well as the overall total heat maximum relative and absolute errors for each option. Figure~\ref{total_heat_sets} compares the average and maximum total heat relative errors for three values of $c_4$; whereas, Figure~\ref{options_timesteps} showcases the temperature distributions for nine solution options at 26s time instance.

It can be observed that the average and maximum errors decrease with the $c_4$ increase, and a close match between the desired total heat temporal variation and the actual total heat variation is easily achieved. Nevertheless, some distinct differences between the temperature distributions are visible in Figure~\ref{options_timesteps}. Furthermore, this figure exemplifies what is mentioned in Section~\ref{methods}, viz., the value of $c_4$ can potentially be used to control the general type or shape of the solution generated. And indeed, for each of the three considered values of $c_4$ the solutions generated from random initial guess sets display similar features. Additionally, it is worth noting that Solution options No. 4 and 6 as well as Solution options No. 7 and 9 look identical in Figure~\ref{options_timesteps}; however, they differ significantly at earlier time instances, whilst adhering to the same solution shape.

\begin{table}[!b]
\small
\begin{threeparttable}[!b]
\caption{Average and maximum relative and absolute total heat errors for nine solution options considered for the stainless steel plate.}
\begin{tabular}{p{0.7cm}|p{2.5cm}p{2.5cm}|p{2.9cm}p{2.9cm}}
  \toprule
Opt. No. & Average relative error [\%] & Maximum relative error [\%] & Average absolute error [W] & Maximum absolute error [W] \\
  \midrule
1 & 1.69 & 9.68 & 0.15 & 0.32 \\
2 & 1.76 & 9.14 & 0.16 & 0.34 \\
3 & 1.64 & 8.33 & 0.15 & 0.30 \\
  \midrule
4 & 2.20 & 10.00 & 0.20 & 0.55 \\
5 & 2.48 & 12.10 & 0.22 & 0.49 \\
6 & 2.19 & 12.35 & 0.19 & 0.41 \\
  \midrule
7 & 0.43 & 2.49 & 0.04 & 0.08 \\
8 & 0.45 & 2.32 & 0.04 & 0.08 \\
9 & 0.42 & 2.37 & 0.04 & 0.08 \\
  \toprule
\end{tabular}
\label{table4_1}
\end{threeparttable}
\end{table}

\subsubsection{Runtimes}\label{runtimes2}

\begin{figure}[!b]
\centering
\includegraphics[width=1.0\textwidth]{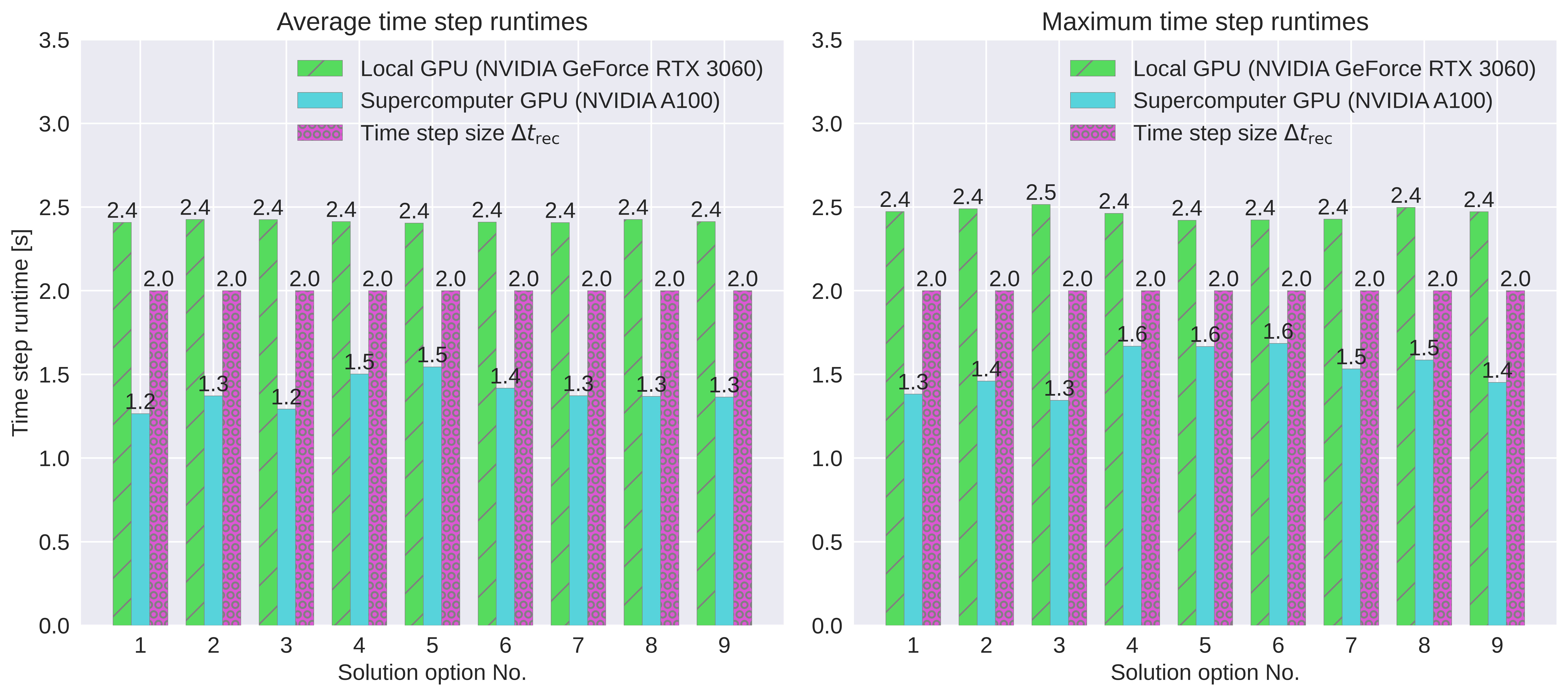}
\caption{Average and maximum time step runtimes for all generated solution options, compared with the time step size $\Delta t_{\mathrm{rec}}$ used; one GPU is used for the local and supercomputer cases.}\label{runtimes_all_options}
\end{figure}

As is required for real time control of an experiment, the simulation can run close to real time. Figure~\ref{runtimes_all_options} compares the average and maximum time step runtimes with the time step size $\Delta t_{\mathrm{rec}}$ of 2s used to generate these solutions. As with the solution reconstruction cases in Sub-section~\ref{ther_field_rec}, the simulation on the local GPU is not quite able to operate in near real time; however, the simulation run on the supercomputer GPU attains excellent runtimes of approximately 1.2s per time step when the time step size is 2s.

\section{Conclusions}\label{concs}

In conclusion, this paper introduces an inverse analysis framework, which relies purely on standard FE discretisations, and highlights its potential by analysing its application to a transient heat conduction in a stainless steel plate, which represents experimental scenario. After the initial FE discretisation the presented workflow deviates significantly from a standard FE workflow used to solve forward problems.

The results in this paper address the problem of inverse solution reconstruction from sparse measurements in near real time, which is directly related to the system monitoring objective (Section~\ref{intro}). Furthermore, they showcase a novel approach for producing multiple options fitting certain requirements in a controlled manner, a functionality essential for the semi-autonomous system control objective (Section~\ref{intro}). This is an innovative way of taking advantage of the intrinsic inverse problem ill-posedness. The demonstrable advantages of the presented methodology include the following:

\begin{enumerate}
    \item It combines the best characteristics of the existing data-driven and physics-based approaches, with or without ML. No copious amounts of training data are required to accomplish acceptable accuracy, while the current runtimes allow for the near real-time operation with a consistent 1-2s offset.
    \item It only makes use of the domain information that might be available in the real experimental setup.
    \item It utilises the traditional industry-standard simulation method for solid materials, which significantly simplifies the process of its integration into the existing simulation workflows in the industry. $\left[ \bm{C} \right]$, $\left[ \bm{K} \right]$, $\left\{ \bm{f} \right\}_h$, and mesh can be generated using any preferred software, while the framework's output can be post-processed and analysed again using any preferred software; and also it can be directly compared with any existing forward simulations. Moreover, this means that all the new FEM-related developments can be potentially employed to enhance its performance; for example, such developments might include novel meshing techniques \cite{meshing}.

\end{enumerate}

While there are still some important considerations to be dealt with, such as temperature-dependency of the material properties and more complex geometries, the current results are encouraging, indicating significant potential for the digital twinning applications.

\section{Acknowledgements}\label{acknow}

This work has been part-funded by the EPSRC Energy Programme [grant number EP/W006839/1]. The authors acknowledge the support of Supercomputing Wales and AccelerateAI projects, which is part-funded by the European Regional Development Fund (ERDF) via the Welsh Government for giving them access to NVIDIA A100 40GB GPUs. Additionally, the authors would like to thank Lloyd Fletcher and Adel Tayeb from UKAEA for the fruitful discussions regarding the physical testing of the heat exchange components and the diagnostic tools used during the experiments.

\section{Declarations}\label{declar}

The source codes along with datasets used in the current paper are available from the corresponding author upon request.



\bibliographystyle{elsarticle-num-names}
\bibliography{bibliography}

\begin{thebibliography}{40}
\expandafter\ifx\csname natexlab\endcsname\relax\def\natexlab#1{#1}\fi
\providecommand{\url}[1]{\texttt{#1}}
\providecommand{\href}[2]{#2}
\providecommand{\path}[1]{#1}
\providecommand{\DOIprefix}{doi:}
\providecommand{\ArXivprefix}{arXiv:}
\providecommand{\URLprefix}{URL: }
\providecommand{\Pubmedprefix}{pmid:}
\providecommand{\doi}[1]{\href{http://dx.doi.org/#1}{\path{#1}}}
\providecommand{\Pubmed}[1]{\href{pmid:#1}{\path{#1}}}
\providecommand{\bibinfo}[2]{#2}
\ifx\xfnm\relax \def\xfnm[#1]{\unskip,\space#1}\fi
%Type = Article
\bibitem[{Yao et~al.(2023)Yao, Yang, Wang, and Zhang}]{DT_review}
\bibinfo{author}{J.-F. Yao}, \bibinfo{author}{Y.~Yang}, \bibinfo{author}{X.-C.
  Wang}, \bibinfo{author}{X.-P. Zhang},
\newblock \bibinfo{title}{Systematic review of digital twin technology and
  applications},
\newblock \bibinfo{journal}{Visual Computing for Industry, Biomedicine, and
  Art} \bibinfo{volume}{6} (\bibinfo{year}{2023}) \bibinfo{pages}{10}.
  \DOIprefix\doi{10.1186/s42492-023-00137-4}.
%Type = Book
\bibitem[{Tarantola(2004)}]{Inverse_problem_theory}
\bibinfo{author}{A.~Tarantola}, \bibinfo{title}{Inverse Problem Theory and
  Methods for Model Parameter Estimation}, \bibinfo{publisher}{Society for
  Industrial and Applied Mathematics}, \bibinfo{year}{2004}.
%Type = Inproceedings
\bibitem[{Li et~al.(2014)Li, Sadigh, Sastry, and Seshia}]{huil1}
\bibinfo{author}{W.~Li}, \bibinfo{author}{D.~Sadigh}, \bibinfo{author}{S.~S.
  Sastry}, \bibinfo{author}{S.~A. Seshia},
\newblock \bibinfo{title}{Synthesis for human-in-the-loop control systems},
\newblock in: \bibinfo{editor}{E.~{\'A}brah{\'a}m},
  \bibinfo{editor}{K.~Havelund} (Eds.), \bibinfo{booktitle}{Tools and
  Algorithms for the Construction and Analysis of Systems},
  \bibinfo{publisher}{Springer Berlin Heidelberg}, \bibinfo{address}{Berlin,
  Heidelberg}, \bibinfo{year}{2014}, pp. \bibinfo{pages}{470--484}.
%Type = Article
\bibitem[{Arridge et~al.(2019)Arridge, Maass, Öktem, and
  Schönlieb}]{Cambridge}
\bibinfo{author}{S.~Arridge}, \bibinfo{author}{P.~Maass},
  \bibinfo{author}{O.~Öktem}, \bibinfo{author}{C.-B. Schönlieb},
\newblock \bibinfo{title}{Solving inverse problems using data-driven models},
\newblock \bibinfo{journal}{Acta Numerica} \bibinfo{volume}{28}
  (\bibinfo{year}{2019}) \bibinfo{pages}{1--174}.
  \DOIprefix\doi{10.1017/S0962492919000059}.
%Type = Article
\bibitem[{Lehký et~al.(2014)Lehký, Keršner, and Novák}]{inverse_param}
\bibinfo{author}{D.~Lehký}, \bibinfo{author}{Z.~Keršner},
  \bibinfo{author}{D.~Novák},
\newblock \bibinfo{title}{Framepid-3pb software for material parameter
  identification using fracture tests and inverse analysis},
\newblock \bibinfo{journal}{Advances in Engineering Software}
  \bibinfo{volume}{72} (\bibinfo{year}{2014}) \bibinfo{pages}{147--154}.
  \DOIprefix\doi{https://doi.org/10.1016/j.advengsoft.2013.10.001},
  \bibinfo{note}{special Issue dedicated to Professor Zdeněk Bittnar on the
  occasion of his Seventieth Birthday: Part 2}.
%Type = Article
\bibitem[{Paruch et~al.(2023)Paruch, Piasecka-Belkhayat, and
  Korczak}]{inverse_param2}
\bibinfo{author}{M.~Paruch}, \bibinfo{author}{A.~Piasecka-Belkhayat},
  \bibinfo{author}{A.~Korczak},
\newblock \bibinfo{title}{Identification of the ultra-short laser parameters
  during irradiation of thin metal films using the interval lattice boltzmann
  method and evolutionary algorithm},
\newblock \bibinfo{journal}{Advances in Engineering Software}
  \bibinfo{volume}{180} (\bibinfo{year}{2023}) \bibinfo{pages}{103456}.
  \DOIprefix\doi{https://doi.org/10.1016/j.advengsoft.2023.103456}.
%Type = Article
\bibitem[{Bangian-Tabrizi and Jaluria(2018)}]{Jaluria}
\bibinfo{author}{A.~Bangian-Tabrizi}, \bibinfo{author}{Y.~Jaluria},
\newblock \bibinfo{title}{An optimization strategy for the inverse solution of
  a convection heat transfer problem},
\newblock \bibinfo{journal}{International Journal of Heat and Mass Transfer}
  \bibinfo{volume}{124} (\bibinfo{year}{2018}) \bibinfo{pages}{1147 – 1155}.
  \DOIprefix\doi{10.1016/j.ijheatmasstransfer.2018.04.053}.
%Type = Article
\bibitem[{Tamaddon-Jahromi et~al.(2020)Tamaddon-Jahromi, Chakshu, Sazonov,
  Evans, Thomas, and Nithiarasu}]{Tamaddon_Jahromi}
\bibinfo{author}{H.~R. Tamaddon-Jahromi}, \bibinfo{author}{N.~K. Chakshu},
  \bibinfo{author}{I.~Sazonov}, \bibinfo{author}{L.~M. Evans},
  \bibinfo{author}{H.~Thomas}, \bibinfo{author}{P.~Nithiarasu},
\newblock \bibinfo{title}{Data-driven inverse modelling through neural network
  (deep learning) and computational heat transfer},
\newblock \bibinfo{journal}{Computer Methods in Applied Mechanics and
  Engineering} \bibinfo{volume}{369} (\bibinfo{year}{2020})
  \bibinfo{pages}{113217}.
  \DOIprefix\doi{https://doi.org/10.1016/j.cma.2020.113217}.
%Type = Article
\bibitem[{Tercan and Meisen(2022)}]{ML_manufacturing}
\bibinfo{author}{H.~Tercan}, \bibinfo{author}{T.~Meisen},
\newblock \bibinfo{title}{Machine learning and deep learning based predictive
  quality in manufacturing: a systematic review},
\newblock \bibinfo{journal}{Journal of Intelligent Manufacturing}
  \bibinfo{volume}{33} (\bibinfo{year}{2022}) \bibinfo{pages}{1879--1905}.
  \DOIprefix\doi{10.1007/s10845-022-01963-8}.
%Type = Article
\bibitem[{Chiu and Lee(2017)}]{ML_manufacturing2}
\bibinfo{author}{H.-W. Chiu}, \bibinfo{author}{C.-H. Lee},
\newblock \bibinfo{title}{Prediction of machining accuracy and surface quality
  for cnc machine tools using data driven approach},
\newblock \bibinfo{journal}{Advances in Engineering Software}
  \bibinfo{volume}{114} (\bibinfo{year}{2017}) \bibinfo{pages}{246--257}.
  \DOIprefix\doi{https://doi.org/10.1016/j.advengsoft.2017.07.008}.
%Type = Article
\bibitem[{{Le Clainche} et~al.(2023){Le Clainche}, Ferrer, Gibson, Cross,
  Parente, and Vinuesa}]{ML_aerospace}
\bibinfo{author}{S.~{Le Clainche}}, \bibinfo{author}{E.~Ferrer},
  \bibinfo{author}{S.~Gibson}, \bibinfo{author}{E.~Cross},
  \bibinfo{author}{A.~Parente}, \bibinfo{author}{R.~Vinuesa},
\newblock \bibinfo{title}{Improving aircraft performance using machine
  learning: A review},
\newblock \bibinfo{journal}{Aerospace Science and Technology}
  \bibinfo{volume}{138} (\bibinfo{year}{2023}) \bibinfo{pages}{108354}.
  \DOIprefix\doi{https://doi.org/10.1016/j.ast.2023.108354}.
%Type = Article
\bibitem[{Szrama and Lodygowski(2024)}]{ML_aerospace2}
\bibinfo{author}{S.~Szrama}, \bibinfo{author}{T.~Lodygowski},
\newblock \bibinfo{title}{Aircraft engine remaining useful life prediction
  using neural networks and real-life engine operational data},
\newblock \bibinfo{journal}{Advances in Engineering Software}
  \bibinfo{volume}{192} (\bibinfo{year}{2024}) \bibinfo{pages}{103645}.
  \DOIprefix\doi{https://doi.org/10.1016/j.advengsoft.2024.103645}.
%Type = Book
\bibitem[{Goodfellow et~al.(2016)Goodfellow, Bengio, and Courville}]{DL}
\bibinfo{author}{I.~Goodfellow}, \bibinfo{author}{Y.~Bengio},
  \bibinfo{author}{A.~Courville}, \bibinfo{title}{Deep Learning},
  \bibinfo{publisher}{The MIT Press}, \bibinfo{year}{2016}.
%Type = Book
\bibitem[{Rasmussen and Williams(2005)}]{GPR}
\bibinfo{author}{C.~Rasmussen}, \bibinfo{author}{C.~Williams},
  \bibinfo{title}{Gaussian Processes for Machine Learning}, Adaptive
  Computation and Machine Learning series, \bibinfo{publisher}{MIT Press},
  \bibinfo{year}{2005}.
%Type = Book
\bibitem[{Mehlig(2021)}]{NNs}
\bibinfo{author}{B.~Mehlig}, \bibinfo{title}{Machine Learning with Neural
  Networks: An Introduction for Scientists and Engineers},
  \bibinfo{publisher}{Cambridge University Press}, \bibinfo{year}{2021}.
  \DOIprefix\doi{10.1017/9781108860604}.
%Type = Article
\bibitem[{Yagawa and Okuda(1996)}]{ANNs}
\bibinfo{author}{G.~Yagawa}, \bibinfo{author}{H.~Okuda},
\newblock \bibinfo{title}{Neural networks in computational mechanics},
\newblock \bibinfo{journal}{Archives of Computational Methods in Engineering}
  \bibinfo{volume}{3} (\bibinfo{year}{1996}) \bibinfo{pages}{435 – 512}.
  \DOIprefix\doi{10.1007/BF02818935}.
%Type = Article
\bibitem[{Hochreiter and Schmidhuber(1997)}]{LSTM}
\bibinfo{author}{S.~Hochreiter}, \bibinfo{author}{J.~Schmidhuber},
\newblock \bibinfo{title}{Long short-term memory},
\newblock \bibinfo{journal}{Neural computation} \bibinfo{volume}{9}
  (\bibinfo{year}{1997}) \bibinfo{pages}{1735--80}.
  \DOIprefix\doi{10.1162/neco.1997.9.8.1735}.
%Type = Article
\bibitem[{Bielajewa et~al.(2024)Bielajewa, Tindall, and
  Nithiarasu}]{bielajewa2023}
\bibinfo{author}{W.~Bielajewa}, \bibinfo{author}{M.~Tindall},
  \bibinfo{author}{P.~Nithiarasu},
\newblock \bibinfo{title}{Comparative study of transformer- and lstm-based
  machine learning methods for transient thermal field reconstruction},
\newblock \bibinfo{journal}{Computational Thermal Sciences: An International
  Journal} \bibinfo{volume}{16} (\bibinfo{year}{2024}).
  \DOIprefix\doi{10.1615/ComputThermalScien.2023049936}.
%Type = Inproceedings
\bibitem[{Vaswani et~al.(2017)Vaswani, Shazeer, Parmar, Uszkoreit, Jones,
  Gomez, Kaiser, and Polosukhin}]{classic_transformer}
\bibinfo{author}{A.~Vaswani}, \bibinfo{author}{N.~Shazeer},
  \bibinfo{author}{N.~Parmar}, \bibinfo{author}{J.~Uszkoreit},
  \bibinfo{author}{L.~Jones}, \bibinfo{author}{A.~N. Gomez},
  \bibinfo{author}{L.~u. Kaiser}, \bibinfo{author}{I.~Polosukhin},
\newblock \bibinfo{title}{Attention is all you need},
\newblock in: \bibinfo{booktitle}{Advances in Neural Information Processing
  Systems}, volume~\bibinfo{volume}{30}, \bibinfo{publisher}{Curran Associates,
  Inc.}, \bibinfo{year}{2017}.
%Type = Phdthesis
\bibitem[{Lewis(2023)}]{GPR_thesis}
\bibinfo{author}{R.~Lewis}, \bibinfo{title}{Simulation driven machine learning
  methods to optimise design of physical experiments and enhance data analysis
  for testing of fusion energy heat exchanger components}, \bibinfo{type}{Phd
  thesis}, Swansea University, \bibinfo{year}{2023}.
%Type = Article
\bibitem[{Zhang et~al.(2023)Zhang, Gong, Zhou, Zhao, Zheng, and
  Yao}]{ZHANG2023106354}
\bibinfo{author}{Y.~Zhang}, \bibinfo{author}{Z.~Gong},
  \bibinfo{author}{W.~Zhou}, \bibinfo{author}{X.~Zhao},
  \bibinfo{author}{X.~Zheng}, \bibinfo{author}{W.~Yao},
\newblock \bibinfo{title}{Multi-fidelity surrogate modeling for temperature
  field prediction using deep convolution neural network},
\newblock \bibinfo{journal}{Engineering Applications of Artificial
  Intelligence} \bibinfo{volume}{123} (\bibinfo{year}{2023})
  \bibinfo{pages}{106354}.
  \DOIprefix\doi{https://doi.org/10.1016/j.engappai.2023.106354}.
%Type = Article
\bibitem[{Zhu et~al.(2023)Zhu, Chen, Ren, and Han}]{ZHU2023120697}
\bibinfo{author}{F.~Zhu}, \bibinfo{author}{J.~Chen}, \bibinfo{author}{D.~Ren},
  \bibinfo{author}{Y.~Han},
\newblock \bibinfo{title}{Transient temperature fields of the tank vehicle with
  various parameters using deep learning method},
\newblock \bibinfo{journal}{Applied Thermal Engineering} \bibinfo{volume}{230}
  (\bibinfo{year}{2023}) \bibinfo{pages}{120697}.
  \DOIprefix\doi{https://doi.org/10.1016/j.applthermaleng.2023.120697}.
%Type = Article
\bibitem[{Raissi et~al.(2019)Raissi, Perdikaris, and Karniadakis}]{PINNs}
\bibinfo{author}{M.~Raissi}, \bibinfo{author}{P.~Perdikaris},
  \bibinfo{author}{G.~E. Karniadakis},
\newblock \bibinfo{title}{Physics-informed neural networks: A deep learning
  framework for solving forward and inverse problems involving nonlinear
  partial differential equations},
\newblock \bibinfo{journal}{Journal of Computational Physics}
  \bibinfo{volume}{378} (\bibinfo{year}{2019}) \bibinfo{pages}{686--707}.
  \DOIprefix\doi{https://doi.org/10.1016/j.jcp.2018.10.045}.
%Type = Article
\bibitem[{Sharma et~al.(2023)Sharma, Evans, Tindall, and Nithiarasu}]{PINNs_P}
\bibinfo{author}{P.~Sharma}, \bibinfo{author}{L.~Evans},
  \bibinfo{author}{M.~Tindall}, \bibinfo{author}{P.~Nithiarasu},
\newblock \bibinfo{title}{Stiff-pdes and physics-informed neural networks},
\newblock \bibinfo{journal}{Archives of Computational Methods in Engineering}
  \bibinfo{volume}{30} (\bibinfo{year}{2023}) \bibinfo{pages}{2929--2958}.
  \DOIprefix\doi{10.1007/s11831-023-09890-4}.
%Type = Article
\bibitem[{Karnakov et~al.(2023)Karnakov, Litvinov, and Koumoutsakos}]{ODIL}
\bibinfo{author}{P.~Karnakov}, \bibinfo{author}{S.~Litvinov},
  \bibinfo{author}{P.~Koumoutsakos},
\newblock \bibinfo{title}{Flow reconstruction by multiresolution optimization
  of a discrete loss with automatic differentiation},
\newblock \bibinfo{journal}{The European Physical Journal E}
  \bibinfo{volume}{46} (\bibinfo{year}{2023}) \bibinfo{pages}{59}.
  \DOIprefix\doi{10.1140/epje/s10189-023-00313-7}.
%Type = Misc
\bibitem[{Balcerak et~al.(2023)Balcerak, Ezhov, Karnakov, Litvinov,
  Koumoutsakos, Weidner, Zhang, Lowengrub, Wiestler, and Menze}]{ODIL_med}
\bibinfo{author}{M.~Balcerak}, \bibinfo{author}{I.~Ezhov},
  \bibinfo{author}{P.~Karnakov}, \bibinfo{author}{S.~Litvinov},
  \bibinfo{author}{P.~Koumoutsakos}, \bibinfo{author}{J.~Weidner},
  \bibinfo{author}{R.~Z. Zhang}, \bibinfo{author}{J.~S. Lowengrub},
  \bibinfo{author}{B.~Wiestler}, \bibinfo{author}{B.~Menze},
  \bibinfo{title}{Individualizing glioma radiotherapy planning by optimization
  of a data and physics informed discrete loss}, \bibinfo{year}{2023}.
  \href{http://arxiv.org/abs/2312.05063}{{\tt arXiv:2312.05063}}.
%Type = Book
\bibitem[{Nithiarasu et~al.(2016)Nithiarasu, Lewis, and
  Seetharamu}]{FEM_Nithiarasu}
\bibinfo{author}{P.~Nithiarasu}, \bibinfo{author}{R.~W. Lewis},
  \bibinfo{author}{K.~N. Seetharamu}, \bibinfo{title}{Fundamentals of the
  Finite Element Method for Heat and Mass Transfer}, \bibinfo{edition}{2} ed.,
  \bibinfo{publisher}{Wiley}, \bibinfo{year}{2016}.
%Type = Misc
\bibitem[{{ANSYS, Inc.}(2024)}]{ansys}
\bibinfo{author}{{ANSYS, Inc.}}, \bibinfo{title}{Ansys: Engineering simulation
  software}, \bibinfo{howpublished}{Available at \url{https://www.ansys.com/}},
  \bibinfo{year}{1970--2024}. \bibinfo{note}{[Software]}.
%Type = Misc
\bibitem[{{{\'E}lectricit{\'e} de France (EDF)}(2024)}]{code_aster}
\bibinfo{author}{{{\'E}lectricit{\'e} de France (EDF)}},
  \bibinfo{title}{{Code\_Aster}: Open source finite element solver, analysis of
  structures and thermomechanics for studies and research},
  \bibinfo{howpublished}{Open source on \url{https://code-aster.org/}},
  \bibinfo{year}{1989--2024}. \bibinfo{note}{[Software]}.
%Type = Book
\bibitem[{Zienkiewicz and Taylor(1989)}]{FEM_Zienkiewicz_V2}
\bibinfo{author}{O.~C. Zienkiewicz}, \bibinfo{author}{R.~L. Taylor},
  \bibinfo{title}{The Finite Element Method: Its Basis and Fundamentals},
  volume~\bibinfo{volume}{2}, \bibinfo{edition}{4} ed.,
  \bibinfo{publisher}{McGraw-Hill Book Company (UK) Limited},
  \bibinfo{year}{1989}.
%Type = Book
\bibitem[{Nocedal and Wright(2006)}]{CG}
\bibinfo{author}{J.~Nocedal}, \bibinfo{author}{S.~J. Wright},
  \bibinfo{title}{Numerical Optimization}, Springer Series in Operations
  Research and Financial Engineering, \bibinfo{edition}{2} ed.,
  \bibinfo{publisher}{Springer New York}, \bibinfo{year}{2006}.
  \DOIprefix\doi{10.1007/978-0-387-40065-5}.
%Type = Book
\bibitem[{Bj{\"o}rck(1996)}]{GN1}
\bibinfo{author}{{\AA}.~Bj{\"o}rck}, \bibinfo{title}{Numerical Methods for
  Least Squares Problems}, \bibinfo{publisher}{Society for Industrial and
  Applied Mathematics}, \bibinfo{year}{1996}.
  \DOIprefix\doi{10.1137/1.9781611971484}.
%Type = Article
\bibitem[{Gratton et~al.(2007)Gratton, Lawless, and Nichols}]{GN2}
\bibinfo{author}{S.~Gratton}, \bibinfo{author}{A.~S. Lawless},
  \bibinfo{author}{N.~K. Nichols},
\newblock \bibinfo{title}{Approximate gauss–newton methods for nonlinear
  least squares problems},
\newblock \bibinfo{journal}{SIAM Journal on Optimization} \bibinfo{volume}{18}
  (\bibinfo{year}{2007}) \bibinfo{pages}{106--132}.
  \DOIprefix\doi{10.1137/050624935}.
%Type = Article
\bibitem[{Baydin et~al.(2017)Baydin, Pearlmutter, Radul, and Siskind}]{AD}
\bibinfo{author}{A.~G. Baydin}, \bibinfo{author}{B.~A. Pearlmutter},
  \bibinfo{author}{A.~A. Radul}, \bibinfo{author}{J.~M. Siskind},
\newblock \bibinfo{title}{Automatic differentiation in machine learning: A
  survey},
\newblock \bibinfo{journal}{J. Mach. Learn. Res.} \bibinfo{volume}{18}
  (\bibinfo{year}{2017}) \bibinfo{pages}{5595–5637}.
%Type = Inbook
\bibitem[{Paszke et~al.(2019)Paszke, Gross, Massa, Lerer, Bradbury, Chanan,
  Killeen, Lin, Gimelshein, Antiga, Desmaison, K\"{o}pf, Yang, DeVito, Raison,
  Tejani, Chilamkurthy, Steiner, Fang, Bai, and Chintala}]{PyTorch}
\bibinfo{author}{A.~Paszke}, \bibinfo{author}{S.~Gross},
  \bibinfo{author}{F.~Massa}, \bibinfo{author}{A.~Lerer},
  \bibinfo{author}{J.~Bradbury}, \bibinfo{author}{G.~Chanan},
  \bibinfo{author}{T.~Killeen}, \bibinfo{author}{Z.~Lin},
  \bibinfo{author}{N.~Gimelshein}, \bibinfo{author}{L.~Antiga},
  \bibinfo{author}{A.~Desmaison}, \bibinfo{author}{A.~K\"{o}pf},
  \bibinfo{author}{E.~Yang}, \bibinfo{author}{Z.~DeVito},
  \bibinfo{author}{M.~Raison}, \bibinfo{author}{A.~Tejani},
  \bibinfo{author}{S.~Chilamkurthy}, \bibinfo{author}{B.~Steiner},
  \bibinfo{author}{L.~Fang}, \bibinfo{author}{J.~Bai},
  \bibinfo{author}{S.~Chintala}, \bibinfo{title}{PyTorch: an imperative style,
  high-performance deep learning library}, \bibinfo{publisher}{Curran
  Associates Inc.}, \bibinfo{address}{Red Hook, NY, USA}, \bibinfo{year}{2019}.
%Type = Article
\bibitem[{Wales and Doye(1997)}]{basinhopping}
\bibinfo{author}{D.~J. Wales}, \bibinfo{author}{J.~P.~K. Doye},
\newblock \bibinfo{title}{Global optimization by basin-hopping and the lowest
  energy structures of lennard-jones clusters containing up to 110 atoms},
\newblock \bibinfo{journal}{The Journal of Physical Chemistry A}
  \bibinfo{volume}{101} (\bibinfo{year}{1997}) \bibinfo{pages}{5111--5116}.
  \DOIprefix\doi{10.1021/jp970984n}.
%Type = Book
\bibitem[{Walpole et~al.(2017)Walpole, Myers, Myers, and Ye}]{probability}
\bibinfo{author}{R.~E. Walpole}, \bibinfo{author}{R.~H. Myers},
  \bibinfo{author}{S.~L. Myers}, \bibinfo{author}{K.~Ye},
  \bibinfo{title}{Probability \& Statistics for Engineers \& Scientists:
  MyStatLab Update}, \bibinfo{edition}{9} ed., \bibinfo{publisher}{Pearson
  Boston}, \bibinfo{address}{Boston}, \bibinfo{year}{2017}.
%Type = Phdthesis
\bibitem[{Hancock(2018)}]{Hancock2018}
\bibinfo{author}{D.~Hancock}, \bibinfo{title}{Employing Additive Manufacturing
  for Fusion High Heat Flux Structures}, \bibinfo{type}{Phd thesis}, University
  of Sheffield, \bibinfo{year}{2018}.
%Type = Article
\bibitem[{Barrett et~al.(2023)Barrett, Bamford, Chuilon, Deighan, Efthymiou,
  Fletcher, Gorley, Grant, Hall, Horsley, Kovari, and Tindall}]{Chimera2}
\bibinfo{author}{T.~Barrett}, \bibinfo{author}{M.~Bamford},
  \bibinfo{author}{B.~Chuilon}, \bibinfo{author}{T.~Deighan},
  \bibinfo{author}{P.~Efthymiou}, \bibinfo{author}{L.~Fletcher},
  \bibinfo{author}{M.~Gorley}, \bibinfo{author}{T.~Grant},
  \bibinfo{author}{T.~Hall}, \bibinfo{author}{D.~Horsley},
  \bibinfo{author}{M.~Kovari}, \bibinfo{author}{M.~Tindall},
\newblock \bibinfo{title}{The chimera facility development programme},
\newblock \bibinfo{journal}{Fusion Engineering and Design}
  \bibinfo{volume}{194} (\bibinfo{year}{2023}) \bibinfo{pages}{113689}.
  \DOIprefix\doi{https://doi.org/10.1016/j.fusengdes.2023.113689}.
%Type = Article
\bibitem[{Zou et~al.(2024)Zou, Lo, Sevilla, Hassan, and Morgan}]{meshing}
\bibinfo{author}{X.~Zou}, \bibinfo{author}{S.~B. Lo},
  \bibinfo{author}{R.~Sevilla}, \bibinfo{author}{O.~Hassan},
  \bibinfo{author}{K.~Morgan},
\newblock \bibinfo{title}{The generation of 3d surface meshes for
  nurbs-enhanced fem},
\newblock \bibinfo{journal}{Computer-Aided Design} \bibinfo{volume}{168}
  (\bibinfo{year}{2024}) \bibinfo{pages}{103653}.
  \DOIprefix\doi{https://doi.org/10.1016/j.cad.2023.103653}.

\end{thebibliography}

% \section{Appendices}\label{secA}
% \renewcommand{\thesubsection}{\Alph{subsection}}
\appendix
\section{1D steady-state example}\label{secA1}

\begin{figure}[!b]
\centering
\includegraphics[width=0.46\textwidth]{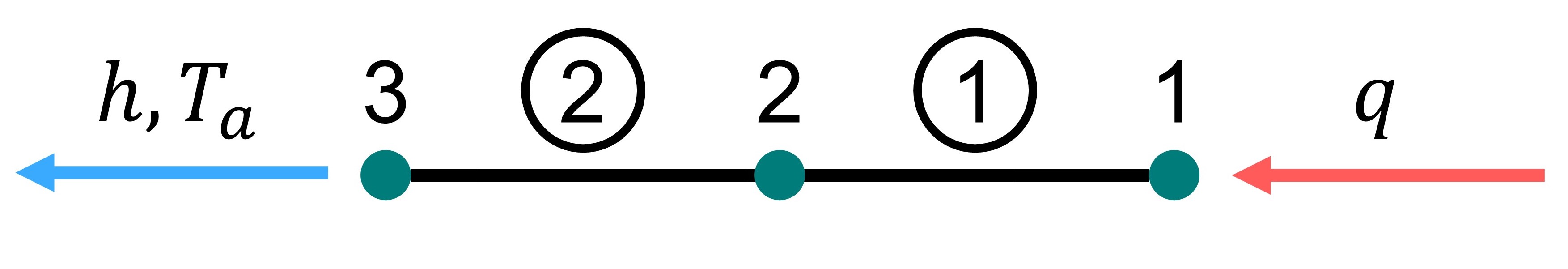}
\caption{The general 1D example with three nodes and two elements.}\label{1D_example}
\end{figure}

This appendix briefly analyses a 1D steady-state example in order to more clearly convey the idea behind the residual loss function term introduced in Section~\ref{methods}. Figure~\ref{1D_example} shows a discretised domain comprised of three nodes and two elements; heat flux $q$ is applied to Node 1, while the domain-atmosphere convection is assumed to take place at Node 3. The specific numbers are not given here as they are not necessary for the understanding of the general concept.

The three equations corresponding to each node are written as follows:

\begin{align}
\text{Node 1:} \quad \left[ \bm{K} \right]_{11} T_1 + \left[ \bm{K} \right]_{12} T_2 + \left[ \bm{K} \right]_{13} T_3 = \left\{ \bm{f} \right\}_{q1} + \left\{ \bm{f} \right\}_{h1} \nonumber \\
\text{Node 2:} \quad \left[ \bm{K} \right]_{21} T_1 + \left[ \bm{K} \right]_{22} T_2 + \left[ \bm{K} \right]_{23} T_3 = \left\{ \bm{f} \right\}_{q2} + \left\{ \bm{f} \right\}_{h2} \nonumber \\
\text{Node 3:} \quad \left[ \bm{K} \right]_{31} T_1 + \left[ \bm{K} \right]_{32} T_2 + \left[ \bm{K} \right]_{33} T_3 = \left\{ \bm{f} \right\}_{q3} + \left\{ \bm{f} \right\}_{h3}
\label{eq23}
\end{align}
Due to the mesh and hence $\left[ \bm{K} \right]$ connectivity and known locations of applied BCs, Eq.~\ref{eq23} can be reduced to the following:
\begin{align}
\text{Node 1:} \quad \left[ \bm{K} \right]_{11} T_1 + \left[ \bm{K} \right]_{12} T_2 = \left\{ \bm{f} \right\}_{q1} \quad\quad\quad \nonumber \\
\text{Node 2:} \quad \left[ \bm{K} \right]_{21} T_1 + \left[ \bm{K} \right]_{22} T_2 + \left[ \bm{K} \right]_{23} T_3 = 0 \; \nonumber \\
\text{Node 3:} \quad  \left[ \bm{K} \right]_{32} T_2 + \left[ \bm{K} \right]_{33} T_3 =  \left\{ \bm{f} \right\}_{h3} \quad\quad\quad
\label{eq24}
\end{align}

Finally, the residual term for this problem can be defined as:
\begin{align}
\text{Residual term} = c_1 \left( \left[ \bm{K} \right]_{21} T_1 + \left[ \bm{K} \right]_{22} T_2 + \left[ \bm{K} \right]_{23} T_3 \right)^2 + \nonumber \\
+ c_1 \left( \left[ \bm{K} \right]_{32} T_2 + \left[ \bm{K} \right]_{33} T_3 - \left\{ \bm{f} \right\}_{h3} \right)^2
\label{eq25}
\end{align}
In Eq.~\ref{eq25}, it can be seen that the equation associated with Node 1 where the unknown heat flux is applied is excluded. Nevertheless, the $\left[ \bm{K} \right]$ matrix connectivity and thus the nodal heat flux balance are preserved, since $T_1$ is still present in the calculation. Node 1 equation is excluded from Eq.~\ref{eq25} as $\left\{ \bm{f} \right\}_{q1}$ is assumed to be unknown; consequently, $\left\{ \bm{f} \right\}_{q1}$ becomes an implicit DoF,  which are not directly adjusted during the optimisation procedure, together with the explicit DoF $\left\{ \bm{T} \right\}$, which are directly adjusted during the optimisation procedure. However, undoubtedly, as DoF, $\left\{ \bm{f} \right\}_{q1}$ and $\left\{ \bm{T} \right\}$ are strongly interdependent.

%\appendix{Appendices}
\section{Weighting coefficient selection for solution reconstruction}\label{secA2}

\begin{figure}[H]
\centering
\includegraphics[width=0.67\textwidth]{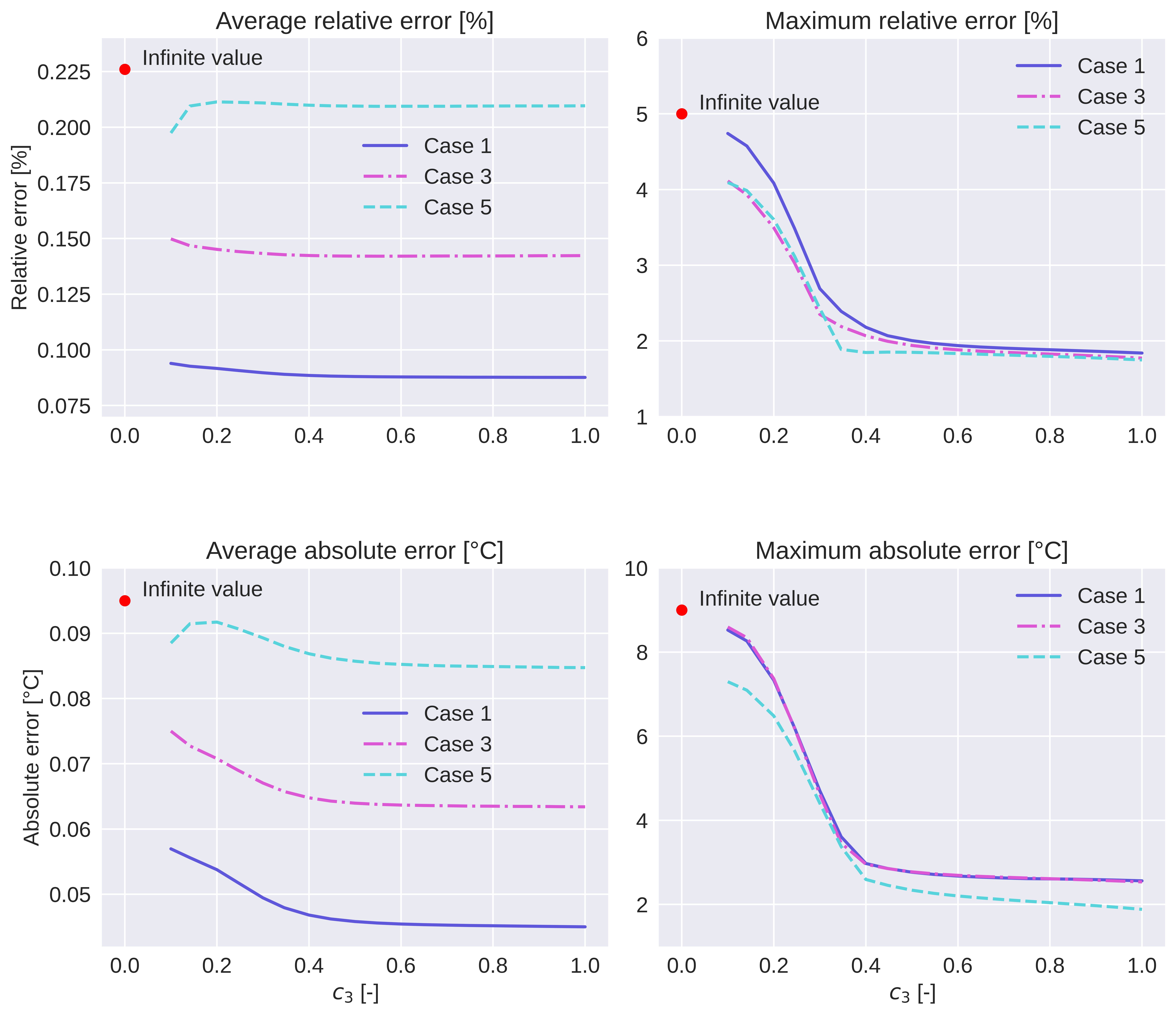}
\caption{Average and maximum relative and absolute solution reconstruction errors (in space and time) for various values of $c_3$ for Cases 1, 3, and 5 (Table~\ref{table2}).}\label{case1-3-5_MULTs}
\end{figure}

\begin{figure}[H]
\centering
\includegraphics[width=0.67\textwidth]{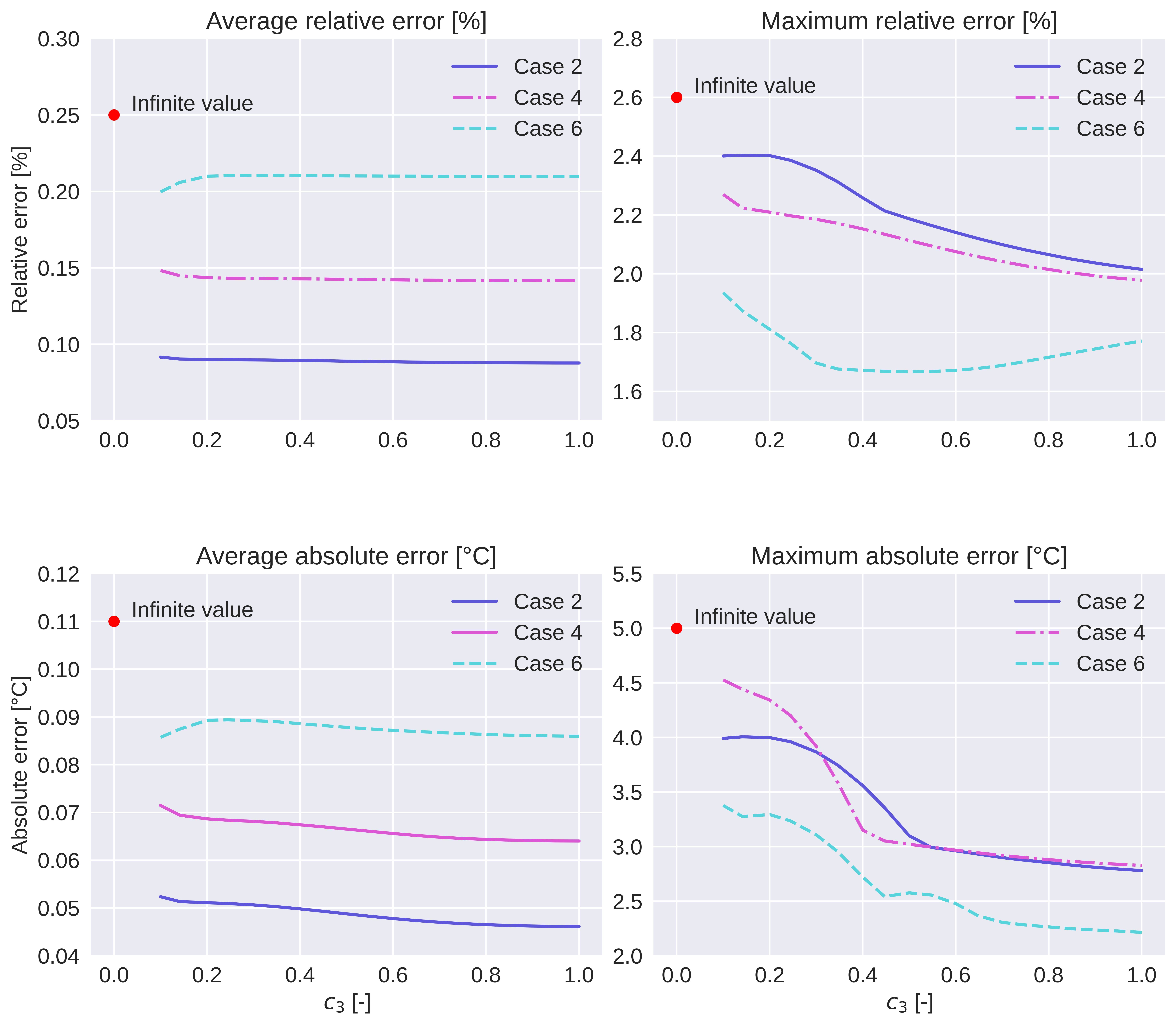}
\caption{Average and maximum relative and absolute solution reconstruction errors (in space and time) for various values of $c_3$ for Cases 2, 4, and 6 (Table~\ref{table2}).}\label{case2-4-6_MULTs}
\end{figure}

The weighting coefficient selection for solution reconstruction is mentioned in Section~\ref{res_disc}; this appendix provides a more detailed view of this process. Figures~\ref{case1-3-5_MULTs} and~\ref{case2-4-6_MULTs} show the average and maximum relative and absolute solution reconstruction errors, in space and time, for various values of $c_3$ for all solution reconstruction cases considered in this paper (Table~\ref{table2}). The cases are split into two groups: 15 measurements (Cases No. 1, 3, and 5 shown in Figure~\ref{case1-3-5_MULTs}), and 9 measurements (Cases No. 2, 4, and 6 shown in Figure~\ref{case2-4-6_MULTs}); nevertheless, a number of common features can be discerned:

\begin{enumerate}
    \item The solutions converge for all values of $c_3$ belonging to $(0, 1]$, and the converged temperature distribution is generally reasonably close to the reference temperature distribution. The solutions diverge rapidly from the reference solution to infinity or to some finite temperature distribution vastly different from the reference temperature distribution for $c_3$ equal to zero, i.e. when no regularisation is applied, thus highlighting the importance of the regularisation term. For every set of measurements there usually exists a temperature distribution perfectly fitting the residual and measurement terms but having an extremely abrupt unphysical temperature variations on the surface; and in the absence of any limiting factor, such as the regularisation term, the optimisation process tends to converge to this unrealistic temperature distribution or not converge at all.
    \item The shapes of the curves are similar between Cases No. 1 and 3, and likewise between Cases No. 2 and 4; the only significant difference is that the average curves are displaced in the direction of $y$ axis. All these cases share the same value of $\Delta t_{\mathrm{rec}}$ that is equal to 1s. However, the values of TSR (Eq.~\ref{eq20}) are different (1.0 for Cases No. 1 and 2, and 10.0 for Cases No. 3 and 4), which might explain the displacement of the average curves in $y$ axis direction.
    \item The shape of the curve for Case No. 5 noticeably differs from the aforementioned curves for Cases No. 1 and 3; the same phenomenon can be observed for Case No. 6 as compared with Cases No. 2 and 4. Examining the various parameter values in Table~\ref{table2} indicates that this change might be caused by the transition of $\Delta t_{\mathrm{rec}}$ from 1s to 2s.
\end{enumerate}

For Cases No. 1-5 the locations of the minima of average relative and maximum relative errors coincide; consequently, the value of 1.0 is selected (Table~\ref{table2}). The situation is is a bit more complex for Case No. 6, since the minima of average relative and maximum relative errors are reached at different values of $c_3$, albeit very close ones. The value of 0.5 is selected for this case as it corresponds to the minimum of the maximum relative error (Table~\ref{table2}).

Overall, the error dependency on $c_3$ seems to be defined by the number of measurements as well as $\Delta t_{\mathrm{rec}}$; consequently, this parameter can be calibrated prior to the experiment. However, it should be noted that usually there are expansive intervals of $c_3$ where the relative errors vary only slightly, for example values of $c_3$ equal or above 0.5.   

\section{Relative and absolute errors for Cases No. 3-6}\label{secA3}

\begin{figure}[!b]
\centering
\includegraphics[width=0.62\textwidth]{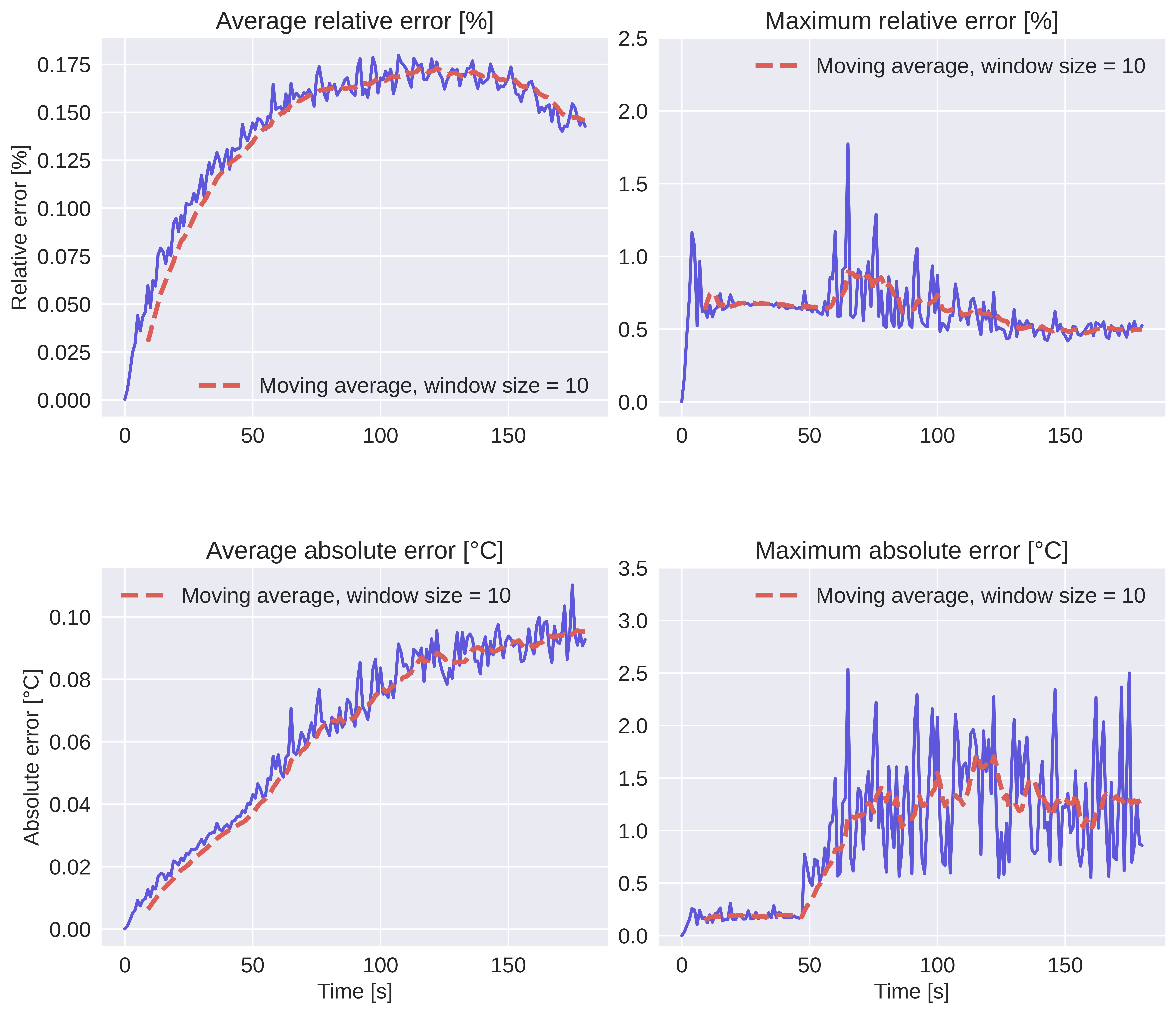}
\caption{The dependence of relative and absolute errors on time for Case No. 3.}\label{case3_errors}
\end{figure}

\begin{figure}[!b]
\centering
\includegraphics[width=0.62\textwidth]{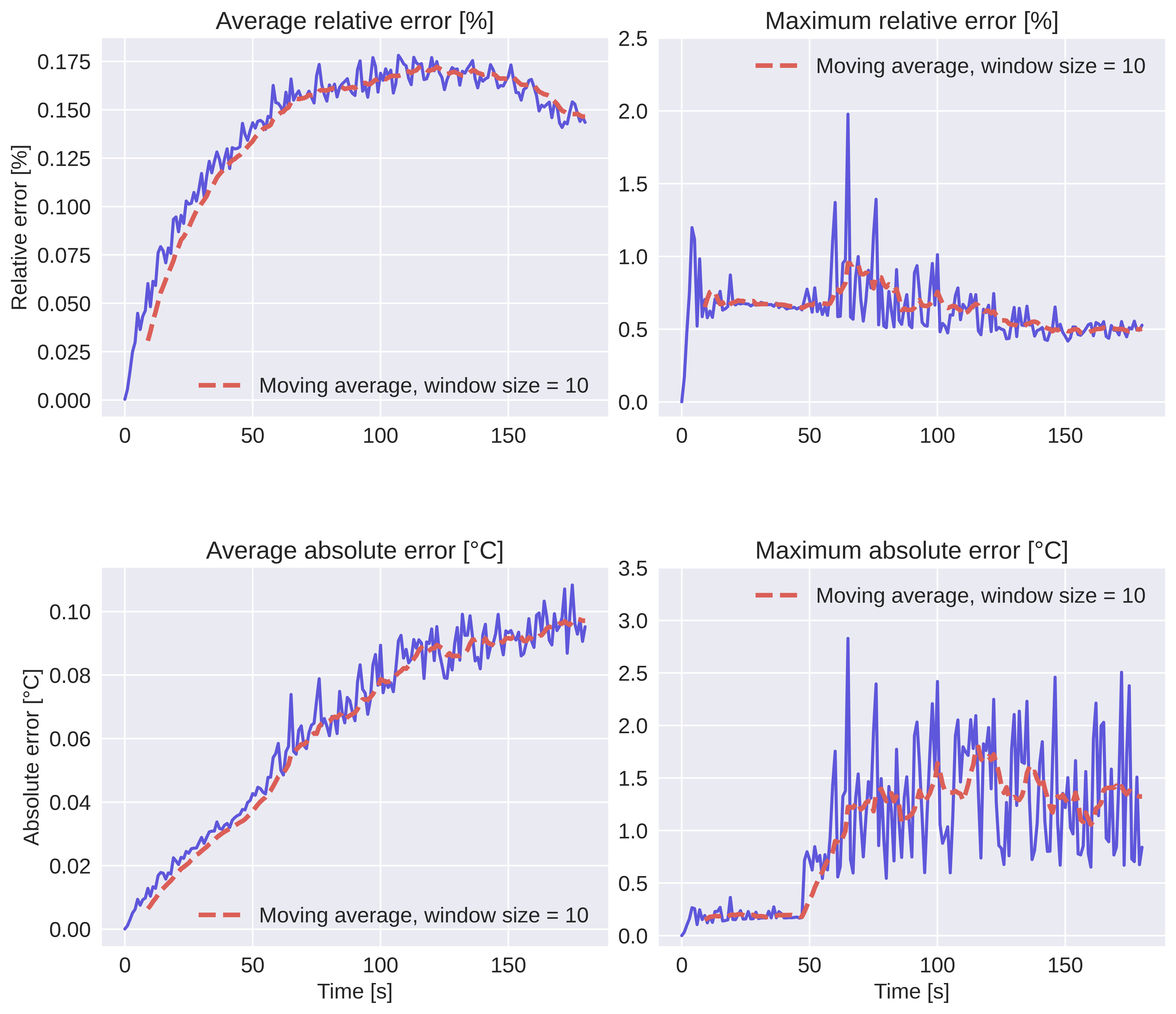}
\caption{The dependence of relative and absolute errors on time for Case No. 4.}\label{case4_errors}
\end{figure}

\begin{figure}[!b]
\centering
\includegraphics[width=0.62\textwidth]{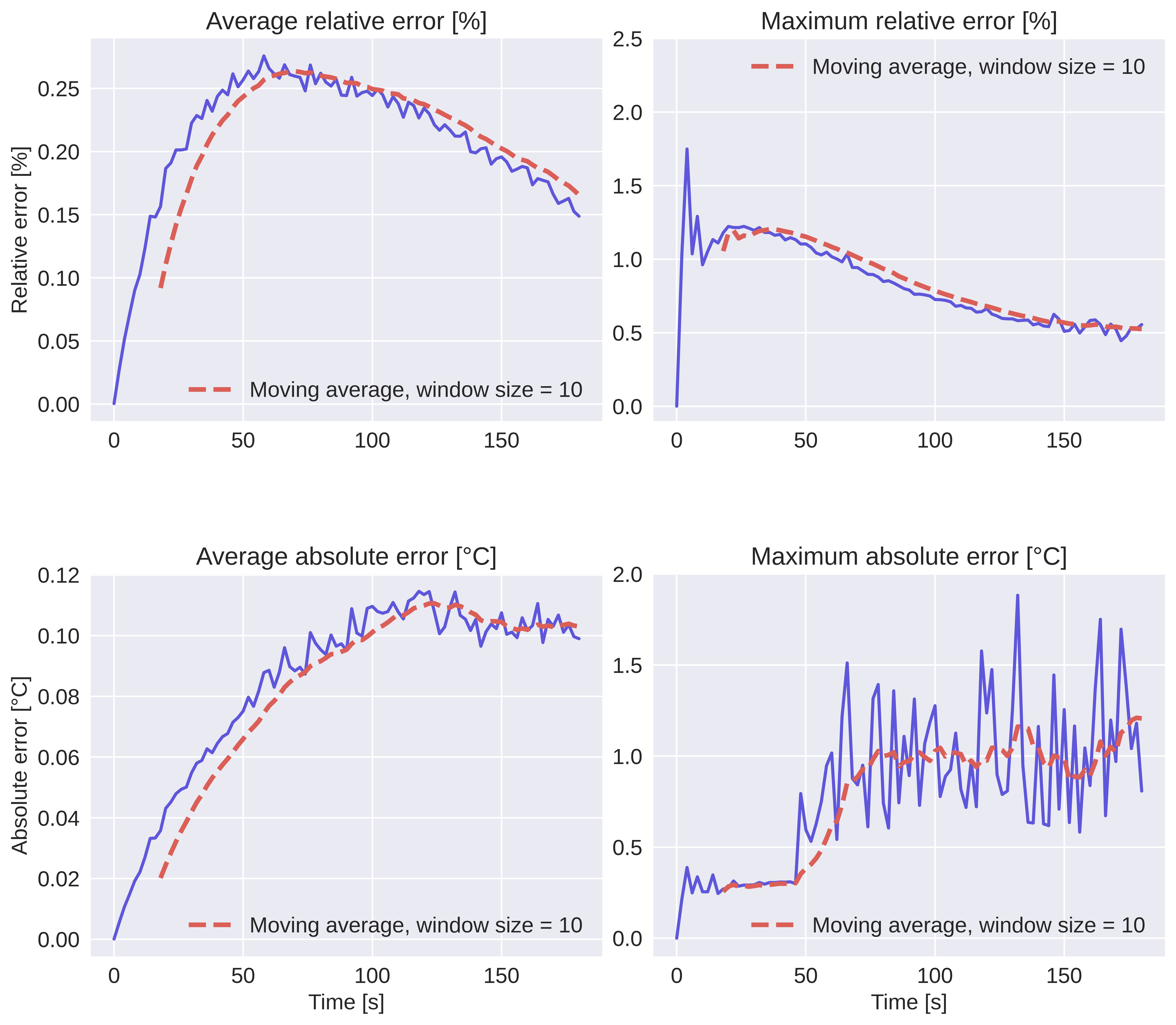}
\caption{The dependence of relative and absolute errors on time for Case No. 5.}\label{case5_errors}
\end{figure}

\begin{figure}[!b]
\centering
\includegraphics[width=0.62\textwidth]{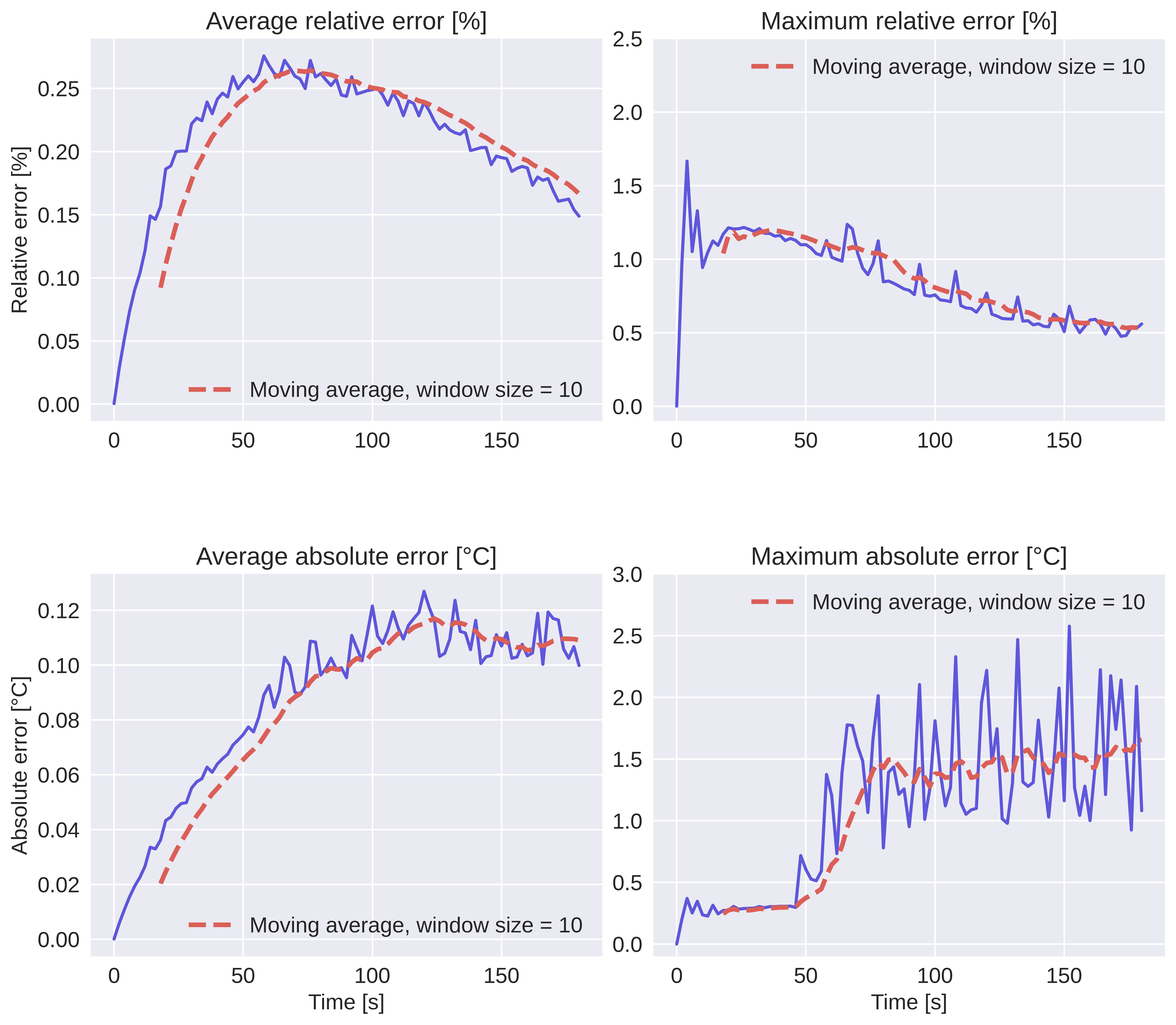}
\caption{The dependence of relative and absolute errors on time for Case No. 6.}\label{case6_errors}
\end{figure}

The figures contained in this appendix,~\Cref{case3_errors,case4_errors,case5_errors,case6_errors}, elaborate on how relative and absolute errors progress over time for Cases No. 3-6.

\end{document}